\documentclass[a4paper,11pt]{article}
\usepackage{amssymb,amsmath,bm}  
\usepackage{graphics,graphicx}   
\usepackage{array,booktabs}      
\usepackage{authblk}  

\usepackage{cite}

\usepackage[margin=2.5cm]{geometry}

\usepackage[linktocpage,
colorlinks = true,
linkcolor = blue,
urlcolor  = blue,
citecolor = red,
anchorcolor = blue]{hyperref}

\usepackage{epsfig}

\numberwithin{equation}{section}

\title{\textbf{Conformal vector dark matter and strongly first-order electroweak phase transition}}
\author[1]{Seyed Yaser Ayazi\thanks{syaser.ayazi@semnan.ac.ir}}
\author[2]{Ahmad Mohamadnejad\thanks{a.mohamadnejad@ut.ac.ir}}
\affil[1]{Physics Department, Semnan University, P.O. Box. 35131-19111, Semnan, Iran}
\affil[2]{Young Researchers and Elite Club, Islamshahr Branch, Islamic Azad University, Islamshahr 3314767653, Iran}
\date{\today}

\begin{document}

\baselineskip 0.6 cm

\maketitle

\begin{abstract}
We study a conformal version of the Standard Model (SM), which apart from SM sector, containing a $ U_{D}(1) $ dark sector with a vector dark matter candidate and a scalar field (scalon). In this model the dark sector couples to the SM sector via a Higgs portal. The theory is scale-invariant in lowest order, therefore the spontaneous symmetry breaking of scale invariance entails the existence of a scalar particle, scalon, with vanishing zeroth-order mass. However, one-loop corrections break scale invariance, so they give mass to the scalon. Because of the scale invariance, our model is subjected to constraints which remove many of the free parameters.  We put constraints to the two remaining parameters from the Higgs searches at the LHC, dark matter relic density and dark matter direct detection limits by PandaX-II. The viable mass region for dark matter is about 1-2 TeV. We also obtain the finite temperature one-loop effective potential of the model and demonstrate that finite temperature effects, for the parameter space constrained by dark matter relic density, induce a strongly first-order electroweak phase transition.
\end{abstract}



\section{Introduction} \label{sec1}
One of the most important challenges of high energy physics is the detection of dark matter (DM) \cite{Bertone:2004pz}. This discovery can explain a number of very important unsolved problems in astrophysics, astronomy and particle physics. One of these unsolved problems is the origin of the spontaneous symmetry breakdown of the electroweak gauge group. In the SM, the electroweak symmetry is broken by Higgs field that has an ad hoc tachyonic mass term. One explanation for the tachyonic mass is radiative symmetry breaking, which is known as Coleman-Weinberg (CW) mechanism \cite{Coleman:1973jx}.

In the CW mechanism spontaneous symmetry breaking is induced at one-loop level from classically scale invariant scalar potential. Scale invariant extensions of SM can address the hierarchy problem which continues to be one of the most crucial questions of modern theoretical physics.
This question that why there is a huge hierarchy in the mass scales of electroweak forces and gravity is related to the naturalness problem.
Systematic cancellation of bosonic and fermionic loop
contributions to the Higgs mass within supersymmetry can also explain the hierarchy problem. However, concerning the null results at the first and second LHC runs \cite{Aaboud:2018zjf,Sirunyan:2018psa}, and other popular theoretical resolutions of the hierarchy problem, such as large extra dimensions, investigating alternative approaches are appealing.

As it was mentioned, one approach of addressing the hierarchy problem is the radical assumption that the fundamental theory describing Nature does not have any scale.
This idea is well worth considering for its potential to be an sparing solution to the hierarchy problem. The CW mechanism with a Higgs does not work for the electroweak symmetry breaking because the large top mass does not permit radiative breaking of the electroweak symmetry, but, simple extensions of the Higgs sector with additional bosonic degrees of freedom are known to be phenomenologically viable (see, e.g., \cite{Foot:2007as,Espinosa:2007qk,Dermisek:2013pta,Antipin:2013exa,Hempfling:1996ht,Tamarit:2014dua,Meissner:2006zh,Foot:2007iy,Iso:2009ss,Iso:2012jn,Englert:2013gz,Abel:2013mya,Das:2015nwk,Hashino:2015nxa,Kubo:2016kpb,Kannike:2016wuy,Ghilencea:2016dsl,Das:2016zue,Arunasalam:2017ajm,Marzola:2017jzl,Chataignier:2018kay,Loebbert:2018xsd}).
On the other hand, the scale-invariant extension of the Higgs sector, is a generic feature of many DM models with scalar \cite{AlexanderNunneley:2010nw,Masina:2013wja,Guo:2014bha,Chang:2007ki,Foot:2010av,Ishiwata:2011aa,Gabrielli:2013hma,Endo:2015nba,
Wang:2015cda,Ghorbani:2015xvz,Plascencia:2015xwa,Helmboldt:2016mpi}, fermionic \cite{Radovcic:2014rea,Altmannshofer:2014vra,Benic:2014aga,Ahriche:2015loa,Ahriche:2016ixu,Oda:2017kwl,YaserAyazi:2018lrv} and vector \cite{Hambye:2013dgv,Carone:2013wla,Khoze:2014xha,Karam:2015jta,Karam:2016rsz,Khoze:2016zfi,Baldes:2018emh} DM candidates.

There are plenty astronomical and cosmological evidences that around 27 percent of the Universe is made of DM.  
According to the dominant paradigm, DM consists of weakly  interacting massive particles (WIMPs) that successfully explain the large scale structures in our Universe.
However, the nature of DM is not well understood, and its particle properties such as spin, mass and interactions all are
unknown. Therefore, it is not surprising that despite many previous models, there are still opportunities for DM model building.

In this paper we consider spin one (vector) gauge fields as DM candidates. Without concerning scale invariance, vector DM \cite{Hambye:2008bq,Hambye:2009fg,DiazCruz:2010dc,Yu:2011by,Farzan:2012hh,Baek:2012se,Baek:2013dwa,Davoudiasl:2013jma,Fraser:2014yga,Graham:2015rva,DiChiara:2015bua,DiFranzo:2015nli,Cembranos:2016ugq,Choi:2017zww,Duch:2017khv,Ahmed:2017dbb,Maru:2018ocf,Chakraborti:2018aae,Belyaev:2018xpf,Saez:2018off} and some of its theoretical and phenomenological aspects such as direct detection \cite{Hisano:2010yh,Baek:2014goa,Yu:2014pra,Chen:2014cbt,Yang:2016zaz,Chen:2016rae,Catena:2018uae,Yepes:2018zkk}, indirect detection \cite{Arina:2009uq,Farzan:2012kk,Choi:2013eua,Ko:2014gha,Farzan:2014foo,Bambhaniya:2016cpr,Chen:2017tva,Yang:2018fje}, and collider physics aspects \cite{Lebedev:2011iq,Duch:2015jta,Chen:2015dea,Kumar:2015wya,Barman:2017yzr} has already been investigated. As it is mentioned, even scale invariant version of vector dark matter models has already been studied. In all the previous scale-invariant models, the dark sector gauge group is non-Abelian \cite{Hambye:2013dgv,Carone:2013wla,Khoze:2014xha,Karam:2015jta,Karam:2016rsz,Khoze:2016zfi,Baldes:2018emh} 
Here we study classically scale-invariant model which apart from SM sector, contains an Abelian $ U_{D}(1) $ dark sector with a vector DM candidate and a scalar field (scalon). In this model the DM couples to the SM sector via a Higgs portal. The physical Higgs will have admixtures of the scalon which can be used to constrain the model's parameter space. The lower limit is set by the LHC constraints
on the mixing angle between the scalon and the Higgs scalar
 \cite{Farzinnia:2013pga,Farzinnia:2014xia}. All the masses in the DM and SM sectors come from a scale generated
dynamically by the CW mechanism. 

After considering relic density and direct detection of DM candidate, we proceed to discuss the finite temperature one-loop corrections to the potential and study electroweak phase transition. Strongly first-order electroweak phase transition is essential for a viable study of baryogenesis which involves investigating
the Sakharov conditions \cite{Sakharov:1967dj}, namely: 1) non-conservation of the baryon
number, 2) violation of C and CP symmetry, and 3) the loss of
thermal equilibrium. Starting from a matter-antimatter
symmetric state, processes obeying the
conditions 1 and 2 are capable of generating a net baryon asymmetry. However, the condition 3 is necessary in order to
hinder the relaxation of such created baryon asymmetry
back to zero. A strongly first-order electroweak phase transition may promote the required departure from thermal
equilibrium for the asymmetry-generating processes.
This condition is satisfied in SM only when the mass of Higgs boson is smaller than 30 GeV \cite{Carrington:1991hz,Anderson:1991zb,Arnold:1992fb,Arnold:1992rz,Dine:1992wr}. Obviously, this range of mass is ruled out after the discovery of a 125 GeV
Higgs boson at the LHC \cite{Chatrchyan:2012xdj,Aad:2012tfa}. However, in extensions of SM including DM candidates it is possible to satisfy the condition for strongly first-order phase transition (For studing electroweak phase transition including a DM candidates see, e.g., \cite{Dimopoulos:1990ai,Chung:2011it,Carena:2011jy,Chowdhury:2011ga,Ahriche:2012ei,Borah:2012pu,Gil:2012ya,Falkowski:2012fb,Cline:2013bln,Ahriche:2013zwa,Fairbairn:2013uta,AbdusSalam:2013eya,Chowdhury:2014tpa,Chao:2014ina,Chao:2017vrq,Gu:2017rzz,Liu:2017gfg,Ghorbani:2017lyk,Shajiee:2018jdq}).

Our model only allows for two independent parameters, the dark gauge coupling and vector DM mass. We have constrained
the model by the observed DM relic abundance as reported by Planck \cite{Aghanim:2018eyx} and WMAP \cite{Hinshaw:2012aka} collaborations. 
We consider LHC constraints on the scalar mixing angle and see that it is satisfied for the parameter space already constrained by DM relic density.
We have also used PandaX-II \cite{Cui:2017nnn} experiment results on the
direct detection of DM to constrain the parameters of the model. Concerning these constraints the mass of the vector DM can be about 1-2 TeV and the upcoming direct detection experiments will be able to sweep a majority of the parameter space. This range of DM mass also implies strongly first-order electroweak phase transition.

Here is the organization of this paper. In section~\ref{sec2} we 
briefly explain the model containing vector DM, 
and we study scale invariance conditions 
for parameters space of the model. Then the thermal 
relic density via freeze-out mechanism is calculated in 
section~\ref{sec3}. DM-nucleon cross section is discussed in
 section~\ref{sec4}. Finite temperature corrections to the effective 
potential is studied in section~\ref{sec5}. In section~\ref{sec6} we 
constrain our model using Planck data for DM relic density and 
PandaX-II direct detection experiment and we demonstrate that electroweak phase transition is strongly first order. Finally, our conclusion comes in section~\ref{sec7}.

\section{The Model} \label{sec2}
We introduce a complex scalar field $ \phi $ which has unit charge under a dark $ U_{D}(1) $ gauge symmetry with a vector field $ V_{\mu} $. Both of these fields are neutral under SM gauge group.
We also consider an additional $ Z_{2} $ symmetry, under which the vector field $ V_{\mu} $ and the scalar field $ \phi $
transform as follows:
\begin{equation}
V_{\mu} \rightarrow - V_{\mu} \, , \quad \phi \rightarrow \phi^{*}, \label{2-1}
\end{equation}
which means in the dark sector we have charge conjugate symmetry.
This discrete symmetry forbids the kinetic mixing between the  the vector field $ V_{\mu} $ and SM $ U_{Y}(1) $ gauge boson $ B_{\mu} $, i.e., $ V_{\mu \nu} B_{\mu \nu} $. Therefore,  the vector field $ V_{\mu} $ is stable and can be considered as a dark matter candidate.
The Lagrangian is given by
\begin{equation}
 {\cal L} ={\cal L}_{SM} + (D_{\mu} \phi)^{*} (D^{\mu} \phi) - V(H,\phi) - \frac{1}{4} V_{\mu \nu} V^{\mu \nu} , \label{2-2}
\end{equation}
where $ {\cal L} _{SM} $ is the SM Lagrangian without the Higgs potential term, $ D_{\mu} \phi = (\partial_{\mu} + i g V_{\mu}) \phi $, $ V_{\mu \nu} = \partial_{\mu} V_{\nu} - \partial_{\nu} V_{\mu}  $, and
the most general scale-invariant potential $ V(H,\phi) $ which is renormalizable and invariant
under gauge and $ Z_{2} $ symmetry is
\begin{equation}
V(H,\phi) = \frac{1}{6} \lambda_{H} (H^{\dagger}H)^{2} + \frac{1}{6} \lambda_{\phi} (\phi^{*}\phi)^{2} + 2 \lambda_{\phi H} (\phi^{*}\phi) (H^{\dagger}H). \label{2-3}
\end{equation}
Note that the quartic portal interaction, $ \lambda_{\phi H} (\phi^{*}\phi) (H^{\dagger}H) $, is the only connection between the dark sector and the SM.

SM Higgs field $ H $ as well as dark scalar $ \phi $ can receive VEVs breaking respectively the electroweak and $ U_{D}(1) $ symmetries.
In unitary gauge, the imaginary component of $ \phi $ can be absorbed as the longitudinal component of $ V_{\mu} $.
In this gauge, we can write
\begin{equation}
H = \frac{1}{\sqrt{2}} \begin{pmatrix}
0 \\ h_{1} \end{pmatrix} \, \, \, and \, \, \, \phi = \frac{1}{\sqrt{2}} h_{2} , \label{2-4}
\end{equation}
where $ h_{1} $ and $ h_{2} $ are real scalar fields which can get VEVs. In this gauge, the tree-level potential becomes
\begin{equation}
V^{tree} = \frac{1}{4 !} \lambda_{H} h_{1}^{4} + \frac{1}{4 !} \lambda_{\phi} h_{2}^{4} + \frac{1}{2} \lambda_{\phi H} h_{1}^{2} h_{2}^{2}. \label{2-5}
\end{equation}
Notice that $ Z_{2} $ symmetry still persists, making $ V_{\mu} $ a stable particle and therefore a DM candidate.

Now consider the Hessian matrix, defined as
\begin{equation}
H_{ij} (h_{1},h_{2}) \equiv \frac{\partial^{2} V^{tree}}{\partial h_{i} \partial h_{j}}. \label{2-6}
\end{equation}
Necessary and sufficient conditions for local minimum of $ V^{tree} $, corresponding to vacuum expectation values $ \langle h_{1} \rangle = \nu_{1} $ $\rm and$ $ \langle h_{2} \rangle = \nu_{2} $, are
\begin{align}
& \frac{\partial V^{tree}}{\partial h_{i}} \bigg\rvert_{\nu_{1},\nu_{2}} = 0 \label{2-7} \\
& \frac{\partial^{2} V^{tree}}{\partial h_{i} ^{2}} \bigg\rvert_{\nu_{1},\nu_{2}} > 0  \label{2-8} \\
& det(H (\nu_{1},\nu_{2})) > 0 , \label{2-9}
\end{align}
where $ det(H (\nu_{1},\nu_{2})) $ is determinant of the Hessian matrix. 
Condition (\ref{2-7}) for non-vanishing VEVs leads to $ \lambda_{H} \lambda_{\phi} = (3! \lambda_{\phi H})^{2} $ and the following constraint
\begin{equation}
\frac{\nu_{1}^{2}}{\nu_{2}^{2}} = - \frac{3! \lambda_{\phi H}}{\lambda_{H}}. \label{2-10}
\end{equation}
Conditions (\ref{2-7}) and (\ref{2-8}) require $ \lambda_{H} > 0 $, $ \lambda_{\phi} > 0 $, and $ \lambda_{\phi H} < 0 $. However, condition (\ref{2-9}) will not be satisfied, because $ det(H (\nu_{1},\nu_{2})) = 0 $. When the determinant of the Hessian matrix is zero, the second derivative test is inconclusive, and the point $ (\nu_{1},\nu_{2}) $ could be any of a minimum, maximum or saddle point. However, in our case, constraint (\ref{2-10}) defines a direction, known as flat direction, in which $ V^{tree} = 0 $. This is the stationary line or a local minimum line. 

Note that in other directions $ V^{tree} > 0 $, and the tree level potential only vanishes along the flat direction, therefore, the full potential of the theory will be dominated by higher-loop contributions along flat direction and specifically by the one-loop effective potential. Adding one-loop effective potential, $ V_{eff}^{1-loop} $, can give a small curvature in the flat direction which picks out a specific value along the ray as the minimum with $ V_{eff}^{1-loop} < 0 $ and vacuum expectation value $ \nu^{2} = \nu_{1}^{2} + \nu_{2}^{2} $ characterized by a RG scale $ \Lambda $. Since at the minimum of the one-loop effective potential $ V^{tree} \geqslant 0 $ and $ V_{eff}^{1-loop} < 0 $, the minimum of $ V_{eff}^{1-loop} $ along the flat direction (where $ V^{tree}=0 $) is a global minimum of the full potential, therefore spontaneous symmetry breaking occurs and we should substitute $ h_{1} \rightarrow \nu_{1} + h_{1} $ and $ h_{2} \rightarrow \nu_{2} + h_{2} $. This breaks the electroweak symmetry with vacuum expectation value $ \nu_{1} = 246 $ GeV.
We first consider the tree level potential. Since $ h_{1} $ and $ h_{2} $ mix with each other, they can be rewritten by the mass eigenstates $ H_{1} $ and $ H_{2} $ as
\begin{equation}
\begin{pmatrix}
H_{1}\\H_{2}\end{pmatrix}
 =\begin{pmatrix} cos \alpha~~~  -sin \alpha \\sin \alpha  ~~~~~cos \alpha
 \end{pmatrix}\begin{pmatrix}
h_1 \\  h_{2}
\end{pmatrix}, \label{2-11}
\end{equation}
where $ H_{2} $ is along the flat direction, thus $ M_{H_{2}} = 0 $, and $ H_{1} $ is perpendicular to the flat direction which we identify it as the SM-like Higgs observed at the LHC with $ M_{H_{1}} = 125 $ GeV. After the symmetry breaking, we have the following constraints:
\begin{align}
& \nu_{2} =  \frac{M_{V}}{g} , \nonumber \\
& sin \alpha =  \frac{\nu_{1}}{\sqrt{\nu_{1}^{2}+\nu_{2}^{2}}} \nonumber \\
& \lambda_{H} =  \frac{3 M_{H_{1}}^{2}}{ \nu_{1}^{2}} cos^{2} \alpha  \nonumber  \\
& \lambda_{\phi} =  \frac{3 M_{H_{1}}^{2}}{ \nu_{2}^{2}} sin^{2} \alpha  \nonumber  \\
& \lambda_{\phi H} =  - \frac{ M_{H_{1}}^{2}}{2 \nu_{1} \nu_{2} } sin \alpha \, cos \alpha , \label{2-12}
\end{align}
where $ M_{V} $ is the mass of vector DM after symmetry breaking. Constraints (\ref{2-12}) severely restrict free parameters of the model up to two independent parameters.
We choose $ M_{V} $ and $ g $ as the independent parameters of the model.

In tree level, the scalon field $ H_{2} $ is massless, and in this case the elastic scattering cross section of DM off nuclei becomes severely large and the model is excluded at once by the DM-nucleon cross section upper bounds provided by direct detection experiments. However, the radiative corrections give a mass to the massless eigenstate $ H_{2} $. Indeed, including the one-loop
corrections to the potential, via the Gildener-Weinberg formalism \cite{Gildener:1976ih}, the scalon mass lifts to
the values that can be even higher than the masses of the other bosons. 

Along the flat direction, the one-loop effective potential, takes the general form \cite{Gildener:1976ih}
\begin{equation}
V_{eff}^{1-loop} = a H_{2}^{4} + b H_{2}^{4} \, \ln \frac{H_{2}^{2}}{\Lambda^{2}}  , \label{2-13}
\end{equation}
where the dimensionless constants $ a $ and $ b $ are given by
\begin{align}
& a =  \frac{1}{64 \pi^{2} \nu^{4}}  \sum_{k=1}^{n} g_{k}  M_{k}^{4} \ln \frac{M_{k}^{2}}{\nu^{2}}  , \nonumber \\
& b = \frac{1}{64 \pi^{2} \nu^{4}} \sum_{k=1}^{n} g_{k}  M_{k}^{4} . \label{2-14}
\end{align}
In (\ref{2-14}), $ M_{k} $ and $ g_{k} $ are, respectively, the  tree-level mass and the internal degrees of freedom of the particle $ k $ (In our convention $ g_{k} $
takes positive values for bosons and negative ones for
fermions).

Minimizing (\ref{2-13}) shows that the potential has a non-trivial stationary point at a value of the RG scale $ \Lambda $, given by
\begin{equation}
\Lambda = \nu \exp \left( \frac{a}{2b} + \frac{1}{4} \right)   , \label{2-15}
\end{equation}
Eq.~(\ref{2-15}) can now be used to find the form of the one-loop effective potential along the flat direction in terms of the one-loop VEV $ \nu $
\begin{equation}
V_{eff}^{1-loop} = b H_{2}^{4} \, \left( \ln \frac{H_{2}^{2}}{\nu^{2}} - \frac{1}{2} \right) , \label{2-16}
\end{equation}
Note that the scalon does not remain massless beyond the tree approximation. Considering $ V_{eff}^{1-loop} $, now the scalon mass will be
\begin{equation}
M_{H_{2}}^{2} = \frac{d^2 V_{eff}^{1-loop}}{d H_{2}^{2}} \bigg\rvert_{\nu} = 8 b \nu^{2} . \label{2-17}
\end{equation}
Regarding (\ref{2-14}), $ M_{H_{2}} $ can be expressed in terms of other particle masses 
\begin{equation}
M_{H_{2}}^{2} = \frac{1}{8 \pi^{2} \nu^{2}} \left( M_{H_{1}}^{4} + 6  M_{W}^{4} + 3  M_{Z}^{4} + 3  M_{V}^{4} - 12 M_{t}^{4}   \right) . \label{2-18}
\end{equation}
where $ M_{W,Z,t} $ being the masses for W and Z gauge bosons, and top quark, respectively. As it was mentioned before $ M_{H_{1}} = 125 $ GeV, $ \nu^{2} = \nu_{1}^{2} + \nu_{2}^{2} $, and $ M_{V} $ is the mass of vector DM.
Notice that in order to $ V_{eff}^{1-loop} $ be a
minimum, it must be less than the value of the potential at the origin, hence it must be negative. From (\ref{2-16}), it is easy to see that this can only happen if $ b > 0 $. On the other hand, considering (\ref{2-18}), one can easily show that in the absence of vector DM mass, $ b < 0 $ or equivalently $ M_{H_{2}}^{2} < 0 $. Therefore, the presence of vector DM is essential in this scenario. Indeed, this constraint, $ M_{H_{2}}^{2} > 0 $, puts a limit on the mass of DM; $ M_{V} > 240 $ GeV.

Note that according to (\ref{2-18}) and (\ref{2-12}), $ M_{H_{2}} $ is completely determined by the independent parameters of the model, i.e., vector DM mass $ M_{V} $ and the coupling $ g $. These constraints are due to the scale invariance conditions which were imposed to the model. In the following sections, we check the validity of our model against DM relic density, and direct detection experimental data.

\section{Relic density via freeze-out} \label{sec3}
Our DM candidate, vector  DM, is a weakly interacting massive particle (WIMP) which is its own antiparticle. In computation of the
relic density of the vector  DM in freeze-out scenario, the standard assumptions are: 1) conservation of the entropy of matter and radiation 2) DM particles were produced thermally, i.e. via interactions with the SM particles in the plasma 3) DM decoupled while the expansion of the Universe was dominated by radiation 4) DM particles were in kinetic and chemical equilibrium before they decoupled.

The current density of vector  DM can be computed by solving the Boltzmann differential equation for the time evolution of vector DM number density $ n_{V} $
\begin{equation}
\dfrac{dn_{V}}{dt} + 3 H n_{V} = - \langle \sigma_{ann} v \rangle (n_{V}^{2} - n_{V,eq}^{2}) . \label{3-1}
\end{equation}
where $ H $ is the Hubble parameter and $ n_{V,eq} $ and $ \langle \sigma_{ann} v \rangle $ are the DM equilibrium number density and the thermally averaged total annihilation cross-section, respectively.
As it was mentioned before, in freeze-out scenario, one of the standard assumptions is the conservation of the entropy of matter and radiation:
\begin{equation}
\dfrac{ds}{dt} + 3 H s = 0 . \label{3-2}
\end{equation}
Here $ s $ is the entropy density. Defining $ Y_{V}=n_{V}/s $ and $ x = M_{V}/T $, with $ T $ the photon temperature, combination of Eq.~(\ref{3-1}) and (\ref{3-2}) gives:
\begin{equation}
\dfrac{dY_{V}}{dx} = \frac{1}{3H} \frac{ds}{dx} \langle \sigma_{ann} v \rangle (Y_{V}^{2} - Y_{V,eq}^{2}) . \label{3-3}
\end{equation}

In standard cosmology, the Hubble parameter is determined
by the mass-energy density $ \rho $ as $ H^{2} = 8 \pi \rho / 3 M_{P}^{2} $ where $ M_{P} = 1.22 \times 10^{19} $ GeV is the Planck mass. On the other hand, the mass-energy density $ \rho $ and entropy density $ s $ are related to the photon temperature by the equations $ \rho = \pi^{2} g_{e} T^{4} / 30 $ and $ s = 2 \pi^{2} h_{e} T^{3} / 45 $, where $ g_{e} $ and $ h_{e} $ are effective degrees of freedom for the energy
density and entropy density, respectively. Regarding these equations, eq.~(\ref{3-3}) can be written as,
\begin{equation}
\dfrac{dY_{V}}{dx} = - \left( \frac{45}{\pi M_{P}^{2}} \right)^{-1/2} \frac{g_{*}^{1/2} M_{V}}{x^{2}} \langle \sigma_{ann} v \rangle (Y_{V}^{2} - Y_{V,eq}^{2}) , \label{3-4}
\end{equation}
where $ g_{*}^{1/2} = \frac{h_{e}}{g_{e}^{1/2}} \left( 1+\frac{T}{3 h_{e}} \frac{d h_{e}}{dT} \right) $.

\begin{figure}[!htb]
\centerline{\hspace{0cm}\epsfig{figure=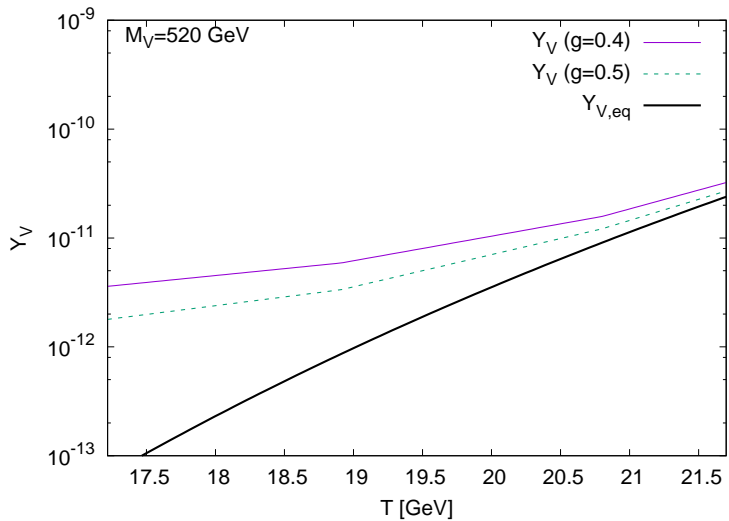,width=10cm}}
\caption{Evolution of $ Y_{V} $ and $ Y_{V,eq} $
during the epoch of DM chemical decoupling (freeze-out).}\label{Y}
\end{figure}

To obtain the present vector  DM abundance $ Y_{V}^{0} $, one should solve differential equation (\ref{3-4}) numerically with the initial condition $ Y_{V} = Y_{V,eq} $
at $ x \simeq 1 $ corresponding to $ T \simeq M_{V} $.

To solve the differential equation (\ref{3-4}), we use {\tt micrOMEGAs} package \cite{Barducci:2016pcb} via {\tt LanHEP} \cite{Semenov:2014rea}. The solution shows that at high temperatures Y closely tracks its equilibrium value $ Y_{V,eq} $. In fact, the interaction rate of vector DM is strong enough to keep them in thermal and chemical equilibrium with the plasma. When the temperature decreases, $ Y_{V,eq} $ becomes exponentially suppressed and $ Y_{V} $ can not reach to its equilibrium value.  But as the temperature decreases, $ Y_{V,eq} $ becomes exponentially suppressed and $ Y_{V} $ is no longer able to track its equilibrium value (see figure~\ref{Y} for an illustration). At the freeze-out temperature, when the vector DM annihilation rate becomes of the order of the Hubble expansion rate, DM production becomes negligible and the WIMP abundance per comoving volume reaches its final value. In figure~\ref{Y}, We have plotted $ Y_{V} $ for two different values of the coupling $ g $. In this figure, freeze-out occurs about $ T_{f}\simeq M_{V}/20 $, where we have chosen $ M_{V}=520 $ GeV. Figure~\ref{Y} illustrates that smaller couplings lead to larger relic densities. This can be understood from the fact that vector DM with larger couplings remain in chemical equilibrium for a longer time, and hence decouple when
the Universe is colder, therefore, its density will be further suppressed.

\begin{figure}[!htb]
\centerline{\hspace{0cm}\epsfig{figure=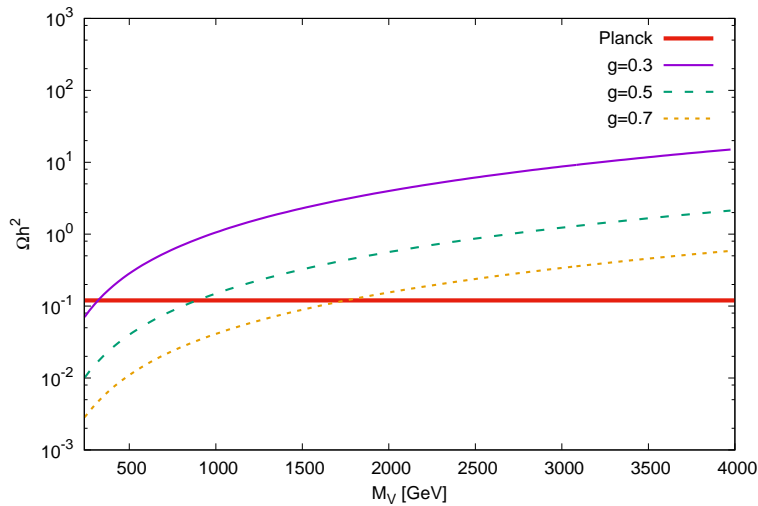,width=10cm}}
\caption{Relic density as a function of vector DM mass for differnet values of coupling $ g $.}\label{R}
\end{figure}

Finally, having $ Y_{V}^{0} $, vector DM relic density can be read as
\begin{equation}
\Omega h^{2} = \frac{\rho_{V}^{0} h^{2}}{\rho_{c}^{0}} = \frac{M_{V} s^{0} Y_{V}^{0} h^{2}}{\rho_{c}^{0}} \simeq 2.755 \times 10^{8} M_{V} Y_{V}^{0} \, \rm GeV^{-1}  , \label{3-5}
\end{equation}
where $ \rho_{c}^{0} $, $ s^{0} $ are the present critical density and entropy density, respectively, and $ h $ is the Hubble constant in units of 100 km/(s.Mpc).
The observational value for DM relic density $ \Omega h^{2} $ is provided by the Planck collaboration \cite{Aghanim:2018eyx} which is
\begin{equation}
\Omega h^{2} = 0.120 \pm 0.001  , \label{3-6}
\end{equation}
In figure~\ref{R}, vector DM relic density versus its mass has been plotted for different values of coupling constant $ g $.
In this figure, larger $ g $ leads to stronger DM-SM interaction which in turn reduces DM relic density.

Now we can compare Eq.~(\ref{3-5}) and (\ref{3-6}) in order to  constrain the parameters space of the model. But first, let us consider another constraint in the next section which arises from DM-nucleon cross section.

\section{Direct detection} \label{sec4}
Direct detection experiments try to detect DM particles through their elastic scattering with nuclei.
These experiments probe the scattering of halo DM particles of highly shielded targets to determine information about their interactions (cross sections) and kinematics (mass). They have explored the parameter space
without finding any evidence of DM. Theoretical and experimental results on direct detection are usually obtained under some simplifying assumptions on the DM profile. In particular, an isothermal profile is often assumed, with $ \rho \propto r^{-2} $ (thus, with a flat rotation curve), a local density of $ \rho_{0} = 0.3 $ $\rm GeVcm^{-3} $, and a Maxwell-Boltzmann velocity distribution with a characteristic velocity of $ v_{0} = 270 ~\rm Km.s^{-1} $.

In this section, we will discuss the discovery potential of the model via direct DM searches. In the present scenario,
at tree level a vector DM particle can collide elastically a nucleon either through $ H_{1} $ exchange
or via $ H_{2} $ exchange, which results in a spin independent cross section \cite{Hambye:2008bq}
\begin{equation}
\sigma_{DM-N} = \frac{4 \lambda_{\phi H}^{2}M_{V}^{2}M_{N}^{2}\mu_{VN}^{2} (M_{H_{1}}^{2} - M_{H_{2}}^{2})^{2}}{\pi M_{H_{1}}^{8} M_{H_{2}}^{4}} f_{N}^{2} , \label{4-1}
\end{equation}
where $ M_{N} $ is the nucleon mass and $ \mu_{VN} = M_{N} M_{V} / (M_{N} + M_{V}) $ is the reduced mass (and $ f_{N} \simeq 0.3 $ parametrizes the Higgs-nucleon coupling).

The best direct detection limits come from the LUX \cite{Akerib:2016vxi}, XENON1T \cite{Aprile:2017iyp}, and PandaX-II \cite{Cui:2017nnn} experiments. Liquid xenon detectors, such as those constructed and operated by the mentioned collaborations, have been leading in detection
capability for heavy-mass WIMPs with masses larger than
10 GeV all the way up to a 100 TeV, which is way beyond the reach of the current generation of colliders. Presently, the PandaX-II \cite{Cui:2017nnn} experiment has set the most stringent upper limit on the spin-independent WIMP-nucleon cross section for a WIMP with mass larger than 100 GeV: 
\begin{eqnarray} \label{constraints}
\rm{PandaX-II}: \sigma_{SI}\leq 8.6\times10^{-47}~cm^2\nonumber 
\end{eqnarray}
Since in our model the mass of the DM is larger than 240 GeV, therefore, we constrain the model with the results of the PandaX-II experiment. It will be seen that this experiment can severely constrain the mass range of the vector DM.

\begin{figure}[!htb]
\centerline{\hspace{0cm}\epsfig{figure=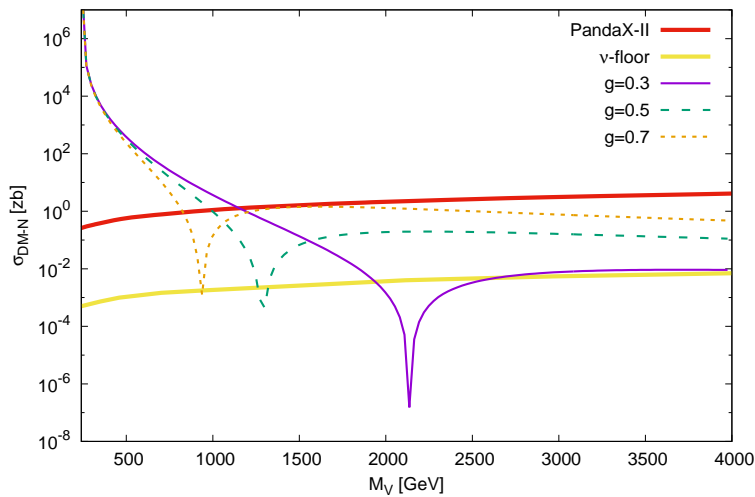,width=10cm}}
\caption{DM-nucleon cross section as a function of vector DM mass for differnet values of coupling $ g $.}\label{D}
\end{figure}

As direct DM experiments go on to enlarge in size, they will become sensitive to the so-called neutrino floor \cite{Billard:2013qya}, i.e., the neutrinos from astrophysical sources, including the Sun, atmosphere, and diffuse supernovae \cite{Cabrera:1984rr,Monroe:2007xp,Strigari:2009bq,Gutlein:2010tq,Harnik:2012ni}. The cross section corresponding to the coherent scattering of neutrinos on nucleons will induce a signal which is similar to the elastic scattering of a WIMP and thus represents an irreducible background \cite{Ruppin:2014bra,Davis:2014ama,Dutta:2015vwa,Dent:2016iht,Ng:2017aur}. Despite possibilities of distinguishing signals from WIMP and neutrino scattering, for example by combining detectors with different target materials, the neutrino floor is usually regarded as the ultimate sensitivity for future Direct Detection experiments such as XENONnT \cite{Aprile:2015uzo}, LZ \cite{Szydagis:2016few} and DARWIN \cite{Aalbers:2016jon}. Therefore, neutrino floor puts a limit on discovery potential of DM.

In figure~\ref{D}, we show the vector DM-nucleon spin-independent elastic scattering cross section, as a function of the vector DM mass for different values of coupling $ g $. Additionally, the upper limit versus WIMP mass for the spin
independent WIMP-nucleon elastic cross sections from the
PandaX-II \cite{Cui:2017nnn} experiment has been depicted. The plot also shows the so-called neutrino floor \cite{Billard:2013qya}, which corresponds to the sensitivity of direct detection experiments to coherent scatterings of neutrinos with nuclei.
According to Eq.~(\ref{4-1}), we expect a dip for the DM-nucleon spin-independent cross section around $ M_{H_{2}} \simeq M_{H_{1}} $. In our model, vector dark matter interacts with nucleon via $ H_{1} $ and $ H_{2} $ mediators. The relevant interaction terms of Lagrangian for $ H_{1} $ mediator are $ \frac{m_{q}}{\nu_{1}} \cos \alpha H_{1} \overline{q} q - \frac{M_{V}^{2}}{\nu_{2}} \sin \alpha H_{1} V_{\mu} V^{\mu} $ and for $ H_{2} $ mediator are $ \frac{m_{q}}{\nu_{1}} \cos \alpha H_{2} \overline{q} q + \frac{M_{V}^{2}}{\nu_{2}} \sin \alpha H_{2} V_{\mu} V^{\mu} $. Therefore, the low-energy 5-dimensional effective interaction terms for DM-quark will be $ \frac{m_{q}}{\nu_{1}} \frac{M_{V}^{2}}{\nu_{2}} \sin \alpha \cos \alpha \left( \frac{1}{M_{H_{2}}^{2}}-\frac{1}{M_{H_{1}}^{2}} \right) \overline{q} q V_{\mu} V^{\mu} $. It means around $ M_{H_{2}} \simeq M_{H_{1}} $ the effective coupling between vector dark matter and quarks goes to zero, leading to a dip in DM-nucleon corss section as it is seen in figure~\ref{D}.

\section{One-loop effective potential at finite temperature} \label{sec5}

In section~\ref{sec1}, it was mentioned that spontaneous symmetry breaking can occur in the one-loop level via Coleman-Weinberg mechanism. However, the symmetry will be restored at high temperature. The character of the symmetry-restoring phase transition is determined by the behavior of the effective potential (free energy) at the critical temperature $ T_{c} $. At this temperature the effective potential has two degenerate minimums.
We will see that the symmetry-restoring at high temperatures is a result of the $ H_{2}^{2} T^{2} $ term that occurs in the one-loop effective potential. This term is the leading-order contribution from the thermal fluctuations of the $ H_{2} $ field. As the temperature rises, the contribution from thermal fluctuations will eventually dominate the one-loop negative (mass-squared) term in the zero-temperature potential and symmetry will be restored. If this phase transition is strongly first order, it can satisfy the condition of departure from thermal equilibrium. This is one
of the three Sakharov conditions \cite{Sakharov:1967dj} necessary for the generation of baryon number asymmetry in the Universe.

At the temperature when the bubbles surrounding the broken phase start to nucleate, one can evade the washout of the baryon number asymmetry by suppression of the baryon number violating interactions induced by electroweak sphalerons \cite{DeSimone:2011ek}.
Sphaleronic interactions are suppressed immediately after the phase transition, which leads to a requirement that $ \nu_{c} $ the vacuum expectation value (VEV) of the
scalon field at the broken phase is larger than the critical temperature, namely
\begin{equation}
\frac{\nu_{c}}{T_{c}} \gtrsim 1 . \label{5-1}
\end{equation}
This is a criteria for strongly electroweak phase transition \cite{Shaposhnikov:1987tw,Shaposhnikov:1986jp}.

In this section, we study conditions of strongly
first-order electroweak phase transition for the model (\ref{5-1}). In section~\ref{sec1}, it is shown that along the flat direction the one-loop potential at zero temperature is given by Eq.~(\ref{2-16}). The finite temperature corrections to this potential at one-loop level can be written as \cite{Carrington:1991hz}
\begin{equation}
V_{T}^{1-loop} (H_{2}) =  \sum_{k=1}^{n} J_{T} (M_{k}(H_{2}),T) + \delta_{kb} D_{T} (M_{k}(H_{2}),\Pi_{k},T) , \label{5-2}
\end{equation}
where $ J_{T} (M_{k}(H_{2}),T) $ is given by \cite{Dolan:1973qd}
\begin{equation}
J_{T} (M_{k}(H_{2}),T) = g_{k} \frac{T^{4}}{2 \pi^{2}} \int_{0}^{\infty} dx x^{2} \ln \left(1 \mp e^{- \sqrt{x^{2}+(M_{k}(H_{2}) / T})^{2}} \right) ,   \label{5-3}
\end{equation}
and $ - $ ($ + $) sign in the integrand corresponds to bosons
(fermions). The tree-level masses $ M_{k} (H_{2}) $ of species $ k $ depend on the scalon field, i.e., $ M_{k}(H_{2}) = \frac{M_{k}}{\nu} H_{2}  $.
The second term in (\ref{5-2}) including the Kronecker delta function $ \delta_{kb} $ takes non-zero value only for bosons and it is given by \cite{Carrington:1991hz}
\begin{equation}
D_{T} (M_{b}(H_{2}),\Pi_{b},T) = g_{b} \frac{T}{12 \pi} \left(M_{b}(H_{2})^{3}-(M_{b}(H_{2})^{2}+\Pi_{b})^{3/2} \right) ,    \label{5-4}
\end{equation}
which is the usual ring improvement (daisy diagram
resummation) for bosonic degrees of freedom depending on the Debye mass $ \sqrt{\Pi_{b}} $ of the boson $ b $.
At leading order, the second contribution of effective potential given by Eq.~(\ref{5-4}) does not depend on scalon field and, neglecting scalon-independent terms, the high-temperature expansion of the thermal integral (\ref{5-3}) leads to
\begin{equation}
V_{T}^{1-loop} (H_{2}) =  c \, T^{2} H_{2}^{2} , \label{5-5}
\end{equation}
where
\begin{equation}
c = \frac{1}{12 \nu^{2}} \sum_{k=1}^{n} c_{k} g_{k} M_{k}^{2} ,  \label{5-6}
\end{equation}
and $ c_{k} = 1 $ ($ c_{k} = - \frac{1}{2} $) for bosons (fermions). Finally, the one-loop effective potential including both one-loop zero temperature (\ref{2-16}) and finite temperature (\ref{5-5}) corrections is given by
\begin{equation}
V_{eff}^{1-loop}(H_{2},T) = b H_{2}^{4} \, \left( \ln \frac{H_{2}^{2}}{\nu^{2}} - \frac{1}{2} \right) + c \, T^{2} H_{2}^{2} ,  \label{5-7}
\end{equation}
where
\begin{align}
b &=  \frac{1}{64 \pi^{2} \nu^{4}}  \left( M_{H_{1}}^{4} + 6  M_{W}^{4} + 3  M_{Z}^{4} + 3  M_{V}^{4} - 12 M_{t}^{4}   \right)  , \label{5-8} \\
c &= \frac{1}{12 \nu^{2}} \left( M_{H_{1}}^{2} + 6  M_{W}^{2} + 3  M_{Z}^{2} + 3  M_{V}^{2} + 6  M_{t}^{2}   \right) . \label{5-9}
\end{align}

\begin{figure}[!htb]
\centerline{\hspace{0cm}\epsfig{figure=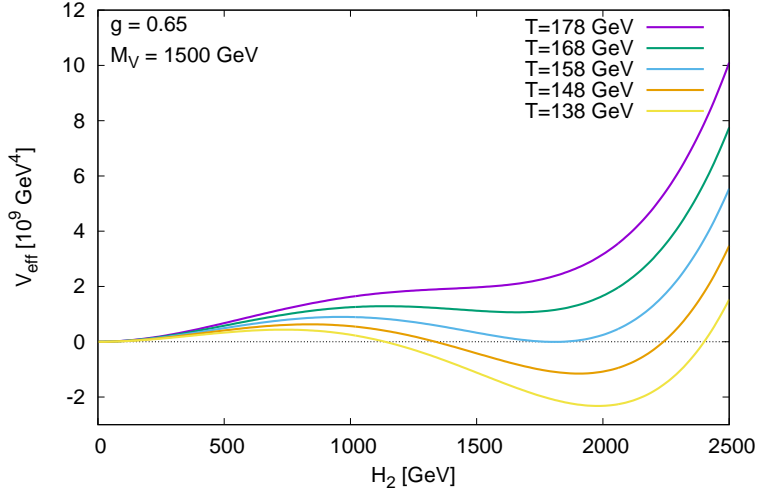,width=10cm}}
\caption{The behavior of the finite temperature one-loop effective potential $ V_{eff}^{1-loop}(H_{2},T) $ for various temperatures. The secondary minimum becomes degenerate with the original one
at a critical temperature $ T_{c} = 158 $ GeV }\label{VH}
\end{figure}

The behavior of the finite temperature one-loop effective potential (\ref{5-7}), for various temperatures, has been depicted in figure~\ref{VH}. In this figure we chose parameters which satisfy both relic density and direct detection constraint. According to this figure, at high temperatures in the early Universe, the global minimum of the potential is located at  $ H_{2} = 0 $. As the Universe expands and temperature decreases, a secondary local minimum begins to appear smoothly, at nonzero values of the field, $ H_{2} \neq 0 $, with a barrier separating the two minimums. The secondary minimum becomes degenerate with the original one
at a critical temperature $ T_{c} $, signaling a first-order electroweak phase transition. At this point the height of the barrier reaches its maximum value. With further temperature drop, the global minimum of the potential will be
located at $ H_{2} \neq 0 $, and the barrier becomes smaller and finally disappearing completely at zero temperature. The phase transition takes place at the critical temperature $ T_{c} $ at which the finite temperature one-loop effective potential (\ref{5-7}) has two degenerate minimums at $ H_{2} = 0 $ and $ H_{2} = \nu_{c} $, i.e.,
\begin{equation}
V_{eff}^{1-loop}(0,T_{c}) = V_{eff}^{1-loop}(\nu_{c},T_{c}) = b \nu_{c}^{4} \, \left( \ln \frac{\nu_{c}^{2}}{\nu^{2}} - \frac{1}{2} \right) + c \, T_{c}^{2} \nu_{c}^{2} = 0 . \label{5-10}
\end{equation}
On the other hand, $ H_{2} = \nu_{c} $ is a local minimum, therefore
\begin{equation}
\frac{\partial V_{eff}^{1-loop}(H_{2},T_{c})}{\partial H_{2}} \bigg\rvert_{H_{2} = \nu_{c}} = 4 b \nu_{c}^{3}  \ln \frac{\nu_{c}^{2}}{\nu^{2}}  + 2 c \, T_{c}^{2} \nu_{c} = 0 . \label{5-11}
\end{equation}
Combining Eqs. (\ref{5-10}) and (\ref{5-11}) together with (\ref{5-1}), the condition for the
electroweak phase transition to be strongly first order becomes
\begin{equation}
\frac{\nu_{c}}{T_{c}} = \sqrt{\frac{c}{b}} \gtrsim 1 . \label{5-12}
\end{equation}

One can also obtain the critical temperature by combining 
Eq.~(\ref{5-10}) or (\ref{5-11}) with (\ref{5-12}) which yields
\begin{equation}
T_{c} = \sqrt{\frac{b}{c}} \nu e^{-\frac{1}{4}} . \label{5-13}
\end{equation}
In the next section, we probe parameter space of the model which simultaneously satisfies constraints from relic density value, direct detection experiment and strongly first-order phase transition.

\section{Results} \label{sec6}

\begin{figure}[!htb]
\begin{center}
\centerline{\hspace{0cm}\epsfig{figure=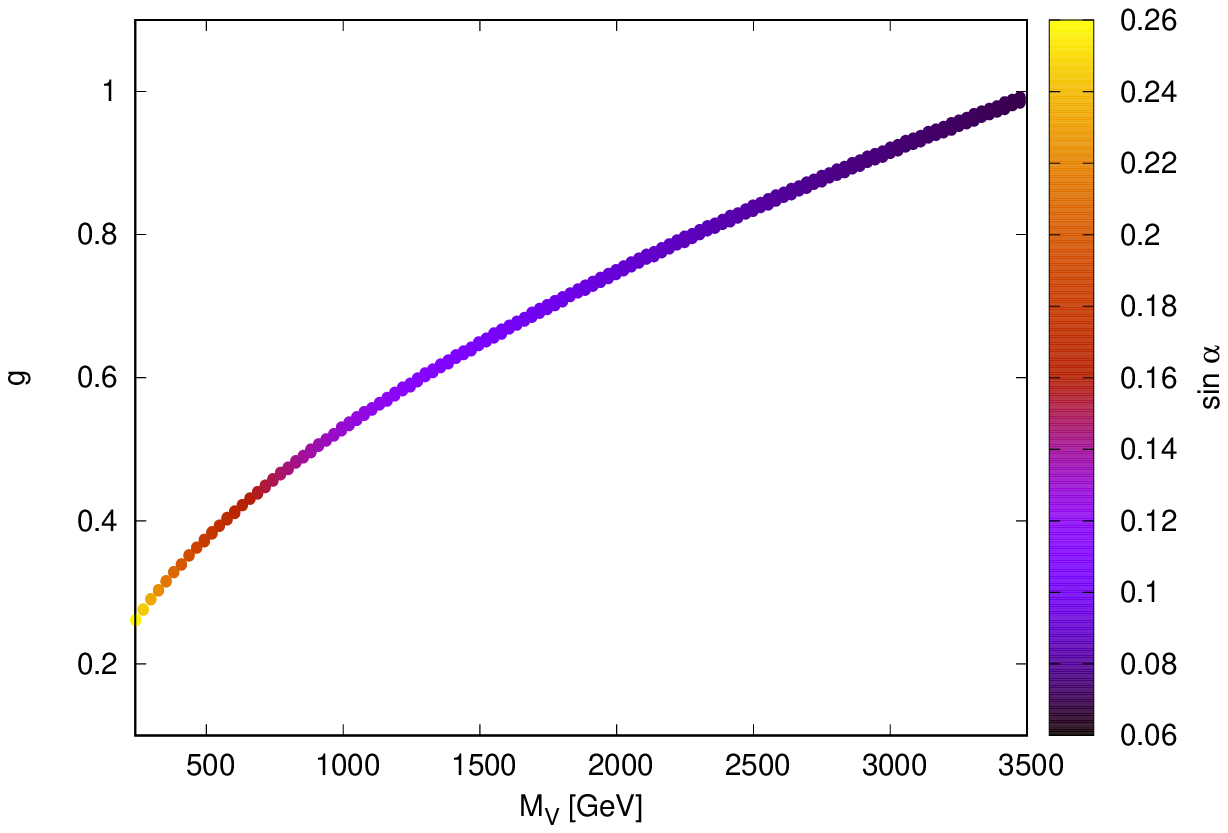,width=7.5cm}\hspace{0cm}\epsfig{figure=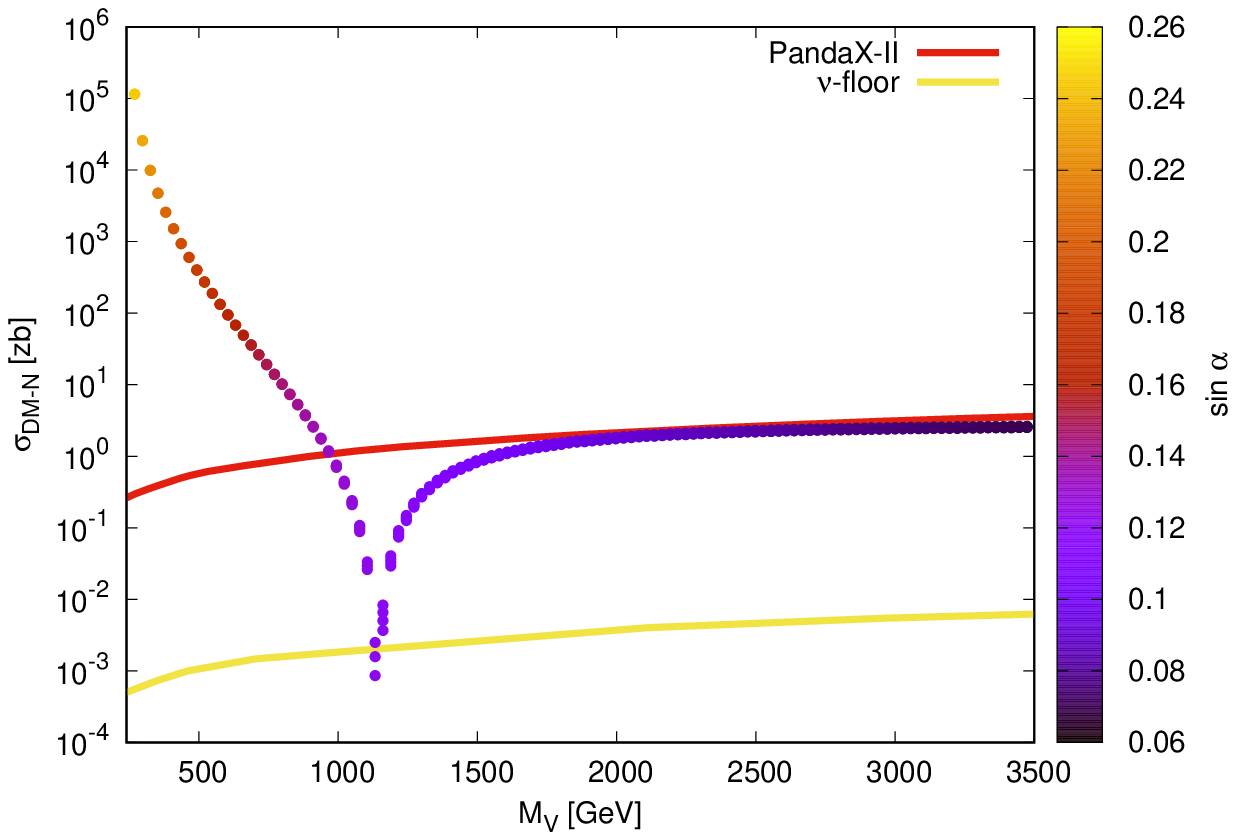,width=7.5cm}}
\centerline{\hspace{0cm}\epsfig{figure=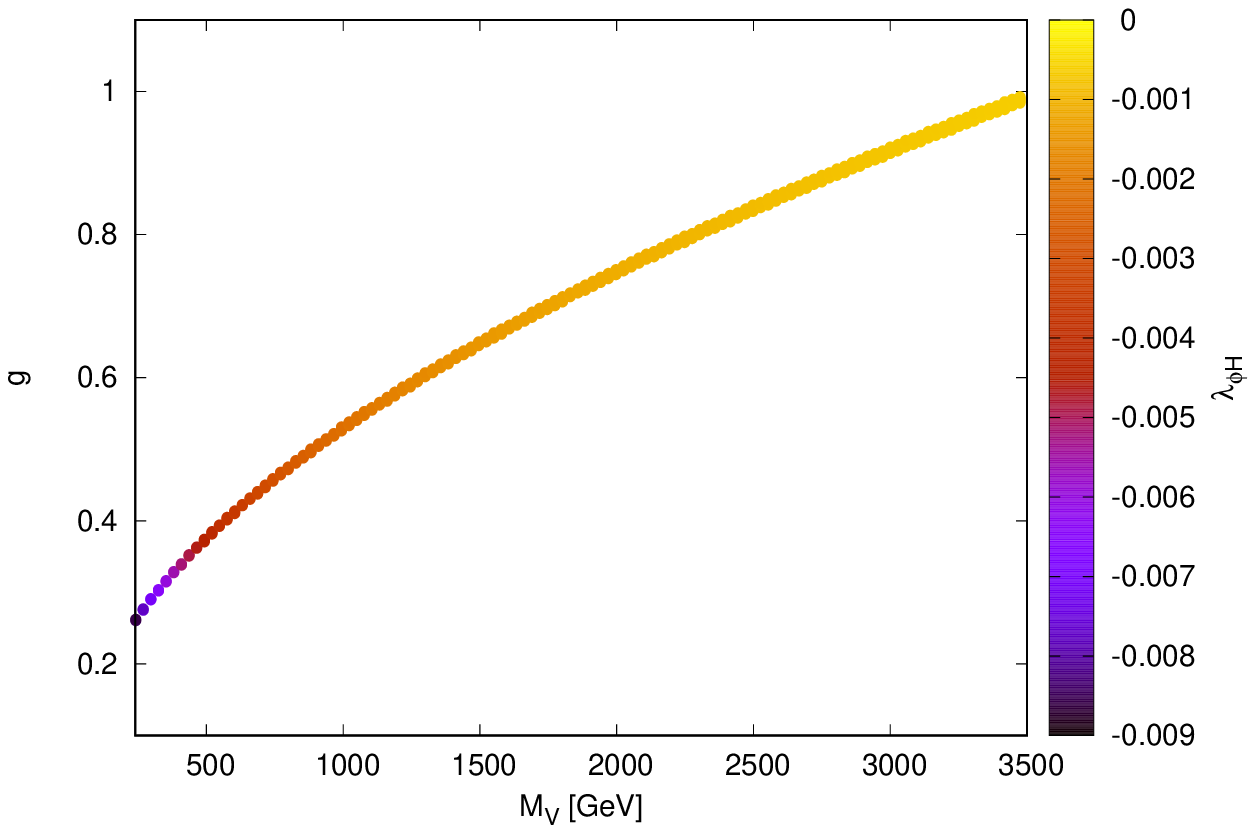,width=7.5cm}\hspace{0cm}\epsfig{figure=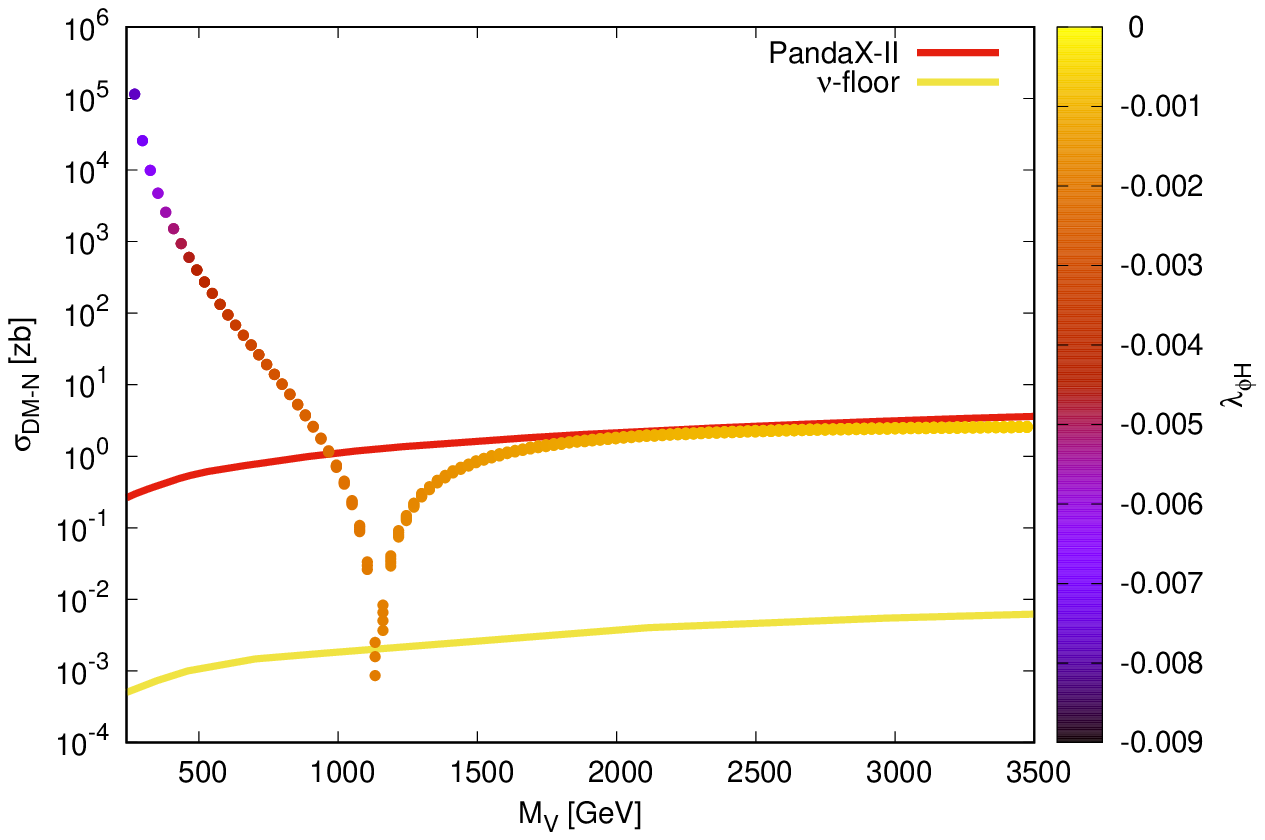,width=7.5cm}}
\centerline{\hspace{0cm}\epsfig{figure=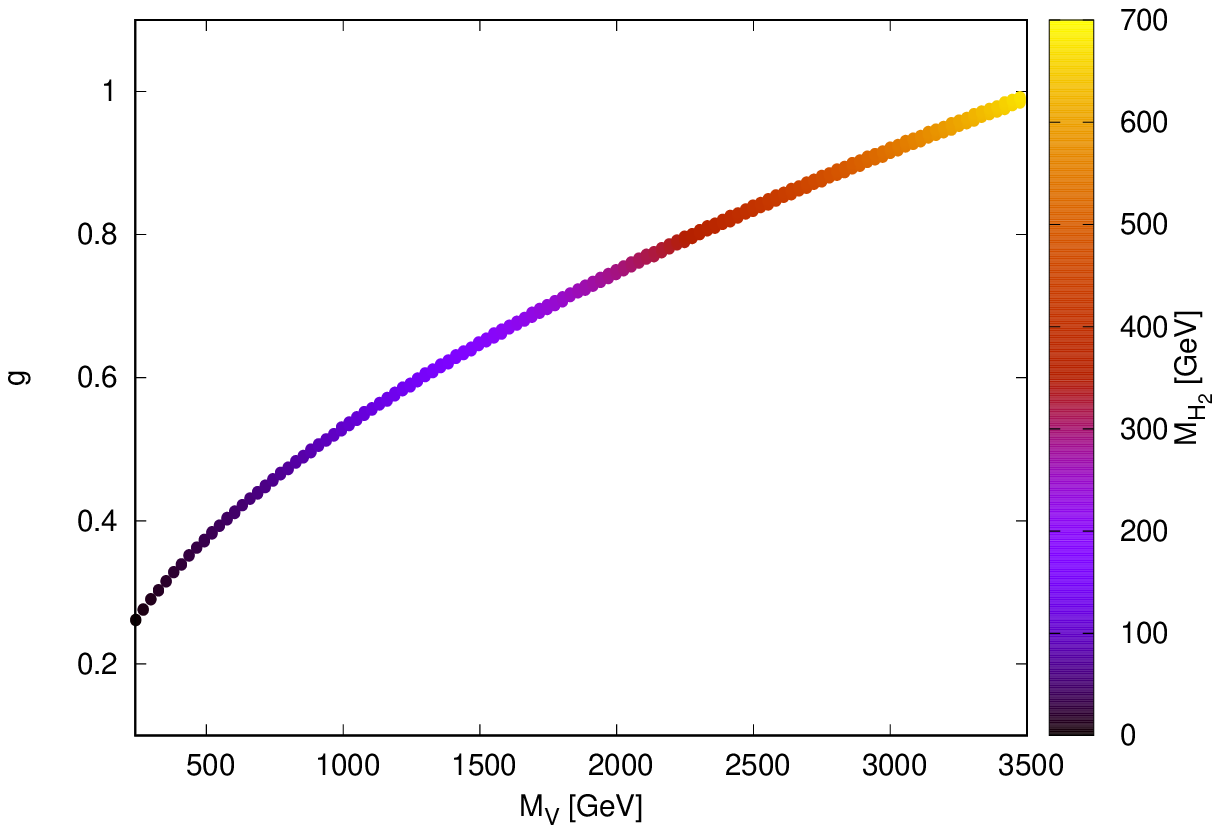,width=7.5cm}\hspace{0cm}\epsfig{figure=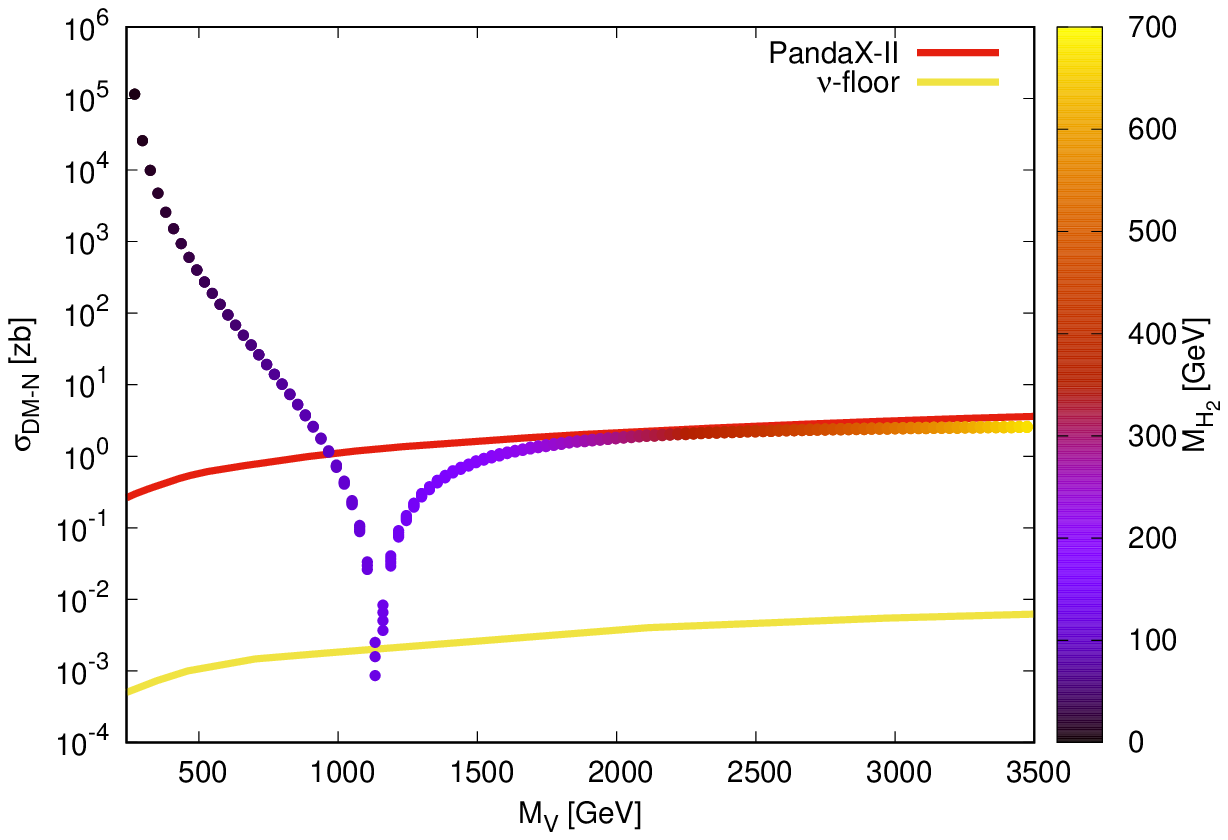,width=7.5cm}}
\centerline{\vspace{-0.7cm}}
\caption{(Left) Ranges of parameters space of the model in $M_V$ and $ g $ plane which are consistent with observed relic density by Planck collaboration. (Right) DM-nucleon cross section as function of DM mass.}\label{constrained Relic Density and Direct Detection}
\end{center}
\end{figure}

In our model, the physical Higgs have admixtures of the scalon which can be used to constrain the parameter space of the model. We can also constrain the two free parameters of the model, i.e., $ M_{V} $ and $ g $, using the Planck data \cite{Aghanim:2018eyx} for DM relic density and PandaX-II \cite{Cui:2017nnn} direct detection experiment. The result has been depicted in figure~\ref{constrained Relic Density and Direct Detection}.
According to this figure the parameter space constrained by relic density is also consistent with the LHC bound on the mixing of the Higgs field to scalon, i.e., $ sin \alpha \leq 0.44 $.

Figure~\ref{constrained Relic Density and Direct Detection} (Left) shows the parameter space which can produce the Planck data for DM relic density. We have assigned three plots showing the values of the three dependent parameters, i.e., $ sin \alpha $, $ \lambda_{\phi H} $, and $ M_{H_{2}} $, via a color bar. In figure~\ref{constrained Relic Density and Direct Detection} (Right), we have calculated DM-nucleon elastic scattering for the parameter space which is already constrained by the relic density bounds required by Planck data. 
Despite the very narrow parameter
space, still for DM masses heavier than around 1 TeV, we have a viable parameter space that respect both the Planck and the PandaX-II bounds.
Note that DM-nucleon cross section for DM masses heavier than 2 TeV lies near PandaX-II upper limit and it will be found or ruled out by the direct detection experiments in the coming years. Given the fact that the bound will be improved greatly, makes the prospects for
discovery very great.

In our model DM annihilation cross section for the parameter space which is already constrained by DM relic density is about $ 2.2 \times 10^{-26} \, cm^{3} / s $. Generally, DM annihilation in the high density regions of the Universe could lead to indirect detection signals, i.e., fluxes
of SM particles, including the flux of continuum gamma rays, positrons, and antiprotons. Searches for all such annihilations products are not yet sensitive enough to reach the typical values of the WIMP cross section for DM masses above 1 TeV \cite{Conrad:2014tla} as found in our model. We conclude that at the moment, our model is not restricted by present DM indirect searches. However, recently the DArk Matter Particle Explorer (DAMPE) has reported an excess in the electron-positron flux of the cosmic rays \cite{Ambrosi:2017wek} which can be interpreted as a signal of the annihilation of DM particle with the mass about 1.5 TeV in a nearby subhalo. For a model-independent analysis of the DAMPE excess due to DM annihilation see \cite{Athron:2017drj}. This feature could also be a statistical fluctuation \cite{Fowlie:2017fya} or may be due to standard astrophysical sources.

DM annihilation in a nearby subhalo with a distance of $ 0.1-0.3 $ kpc can explain the DAMPE peak for the annihilation cross section about $ 2-4 \times 10^{-26} \, cm^{3} / s $ and the DM density about 17-35 times greater than the local density of DM \cite{Yuan:2017ysv}.
In our model, vector DM mass around 1.5 TeV can pass relic density and direct detection constraints and its annihilation cross section is about $ 2.2 \times 10^{-26} \, cm^{3} / s $ which is large enough that it might account for the DAMPE peak.
However, in order to explain DAMPE data it is necessary to generalize the model such that DM annihilation through $ e^{+} e^{-} $ channel be dominated. Therefore, we anticipate that including leptophilic interactions, $ \sum\limits_{l=e,\mu,\tau} g_{l} \overline{l} H_{2} l $, to the model might also explain DAMPE excess.

\begin{figure}[!htb]
\centerline{\hspace{0cm}\epsfig{figure=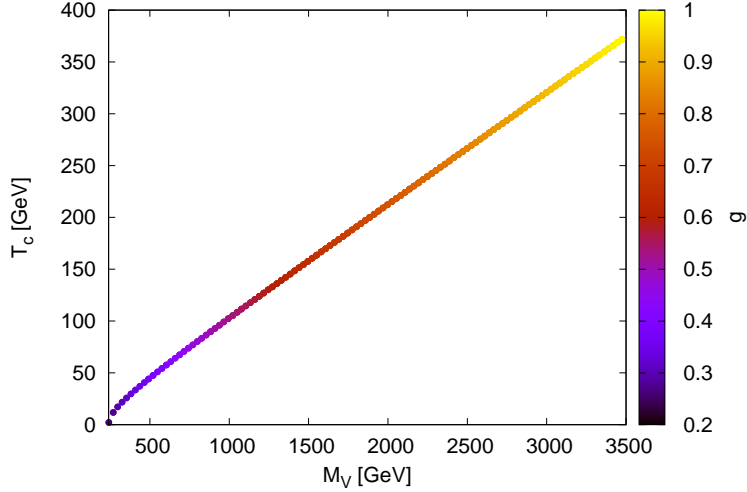,width=10cm}}
\caption{Critical temperature versus mass of vector DM for the parametrs which satisfy DM relic density constraint.}\label{Tc}
\end{figure}

\begin{figure}[!htb]
\centerline{\hspace{0cm}\epsfig{figure=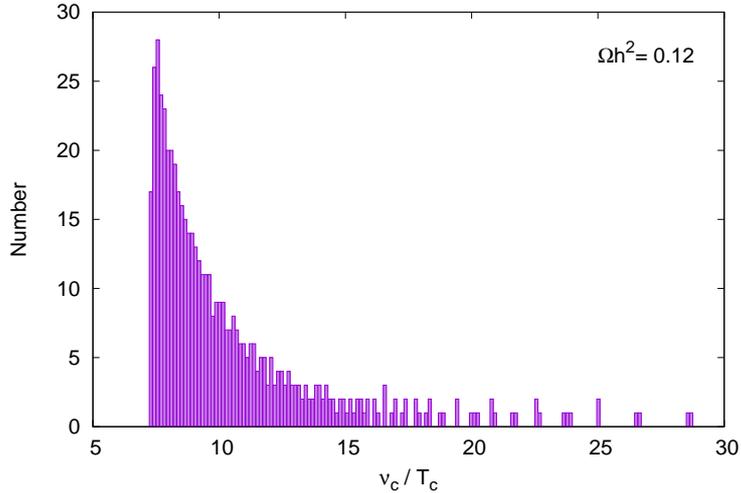,width=10cm}}
\caption{The frequency distribution of the order parameter $ \nu_{c} / T_{c} $ for the samples which are satisfying DM relic density constraint. This figure shows that $ \nu_{c} / T_{c} > 7 $, implying a strongly first order phase transition.}\label{Histogram}
\end{figure}

Finally, in figure~\ref{Tc} we have depicted the critical temperature in the first-order electroweak phase transition as a function of parameters of the model, $ g $ and $ M_{V} $, which already satisfied the DM relic density constraint. It is seen that viable range for the mass of DM constrained by relic density and direct detection is about 1-2 TeV. According to figure~\ref{Tc}, this range implies that $ T_{c} $ is about 100-200 GeV. 
In the end, we have depicted the distribution of the order parameter $ \nu_{c} / T_{c} $ in figure~\ref{Histogram} for the samples which are satisfying DM relic density constraint. According to this figure $ \nu_{c} / T_{c} > 7 $ implying a strongly first order electroweak phase transition which can address electroweak baryogenesis.

\section{Conclusion} \label{sec7}

In this paper, we studied a simple conformal extension of the SM in which radiative symmetry breaking within the Coleman-Weinberg mechanism can take place. Conformal extensions of the SM are a possible solution to the hierarchy problem through the dynamical generation of mass scales.
Here we proposed a minimal classically scale invariant
version of the SM, enlarged by a dark $ U_{D}(1) $
gauge group which incorporates a vector boson (vector DM) and a
scalar field (scalon). The dark sector was radiatively broken
through the Coleman-Weinberg mechanism and a mass
scale was communicated to the electroweak
through the portal interactions of the scalon
with the Higgs field. We obtained one-loop scalar potential employing Gildener-Weinberg formalism and observed that scalon mass, although zero at tree level, can receive large quantum corrections. Due to scale invariance, the model has only two independent parameters.

After setting up the model, we proceed to calculate relic density by solving the Boltzmann equation, and then we obtained  
DM-nucleon cross section. We also studied finite temperature effects and obtained one-loop effective potential at high temperatures in order to investigate electroweak phase transition. 
In order to prevent the
washout of the matter-antimatter asymmetry, strongly first-order electroweak phase transition is a
necessary condition for the successful implementation of
electroweak baryogenesis. Matching calculated relic density and DM-nucleon cross section to the observed values coming from Planck and PandaX-II experiments, respectively, we constrained the two independent parameters of the model.
It is shown that a part of the parameter space of the model will be excluded and the rest of the parameter space is within the reach of the future direct detection experiments. It has shown that the parameter space constrained by relic density demonstrate strongly first-order electroweak phase transition. Considering constraint coming from PandaX-II direct detection experiment, we obtained viable mass range for DM which is about 1-2 TeV. This range puts a limit on critical temperature, $ T_{c} $= 100-200 GeV.
The model is also compatible with the experimental bound on the mixing of Higgs field to other scalar which is given
by $ sin \alpha \leq 0.44 $.

Conformal extension of the SM that we considered here, predicts new scalar boson (scalon), and vector DM with a definite mass range that can be discovered by future colliders and probed by upcoming direct detection experiments.

\section*{Acknowledgments}
The work of A.M. is supported financially by the Young Researchers and Elite Club of Islamshahr Branch of Islamic Azad University.


\begin{thebibliography}{99}

\bibitem{Bertone:2004pz} 
  G.~Bertone, D.~Hooper and J.~Silk,
  Particle dark matter: Evidence, candidates and constraints,
  Phys.\ Rept.\  {\bf 405}, 279 (2005)
  [hep-ph/0404175].


\bibitem{Coleman:1973jx} 
  S.~R.~Coleman and E.~J.~Weinberg,
  Radiative Corrections as the Origin of Spontaneous Symmetry Breaking,
  Phys.\ Rev.\ D {\bf 7}, 1888 (1973).


\bibitem{Aaboud:2018zjf} 
  M.~Aaboud {\it et al.} [ATLAS Collaboration],
  Search for supersymmetry in final states with charm jets and missing transverse momentum in 13 TeV $pp$ collisions with the ATLAS detector,
  JHEP {\bf 1809}, 050 (2018)
  [arXiv:1805.01649 [hep-ex]].


\bibitem{Sirunyan:2018psa} 
  A.~M.~Sirunyan {\it et al.} [CMS Collaboration],
  Search for supersymmetry in events with a photon, a lepton, and missing transverse momentum in proton-proton collisions at $\sqrt{s} =$ 13 TeV,
  JHEP {\bf 1901}, 154 (2019)
  [arXiv:1812.04066 [hep-ex]].


\bibitem{Foot:2007as} 
  R.~Foot, A.~Kobakhidze and R.~R.~Volkas,
  Electroweak Higgs as a pseudo-Goldstone boson of broken scale invariance,
  Phys.\ Lett.\ B {\bf 655}, 156 (2007)
  [arXiv:0704.1165 [hep-ph]].


\bibitem{Espinosa:2007qk} 
  J.~R.~Espinosa and M.~Quiros,
  Novel Effects in Electroweak Breaking from a Hidden Sector,
  Phys.\ Rev.\ D {\bf 76}, 076004 (2007)
  [hep-ph/0701145].


\bibitem{Dermisek:2013pta} 
  D.~Chway, T.~H.~Jung, H.~D.~Kim and R.~Dermisek,
  Radiative Electroweak Symmetry Breaking Model Perturbative All the Way to the Planck Scale,
  Phys.\ Rev.\ Lett.\  {\bf 113}, no. 5, 051801 (2014)
  [arXiv:1308.0891 [hep-ph]].


\bibitem{Antipin:2013exa} 
  O.~Antipin, M.~Mojaza and F.~Sannino,
  Conformal Extensions of the Standard Model with Veltman Conditions,
  Phys.\ Rev.\ D {\bf 89}, no. 8, 085015 (2014)
  [arXiv:1310.0957 [hep-ph]].


\bibitem{Hempfling:1996ht} 
  R.~Hempfling,
  The Next-to-minimal Coleman-Weinberg model,
  Phys.\ Lett.\ B {\bf 379}, 153 (1996)
  [hep-ph/9604278].


\bibitem{Tamarit:2014dua} 
  C.~Tamarit,
  Higgs vacua with potential barriers,
  Phys.\ Rev.\ D {\bf 90}, no. 5, 055024 (2014)
  [arXiv:1404.7673 [hep-ph]].


\bibitem{Meissner:2006zh} 
  K.~A.~Meissner and H.~Nicolai,
  Conformal Symmetry and the Standard Model,
  Phys.\ Lett.\ B {\bf 648}, 312 (2007)
  [hep-th/0612165].


\bibitem{Foot:2007iy} 
  R.~Foot, A.~Kobakhidze, K.~L.~McDonald and R.~R.~Volkas,
  A Solution to the hierarchy problem from an almost decoupled hidden sector within a classically scale invariant theory,
  Phys.\ Rev.\ D {\bf 77}, 035006 (2008)
  [arXiv:0709.2750 [hep-ph]].


\bibitem{Iso:2009ss} 
  S.~Iso, N.~Okada and Y.~Orikasa,
  Classically conformal $B^-$ L extended Standard Model,
  Phys.\ Lett.\ B {\bf 676}, 81 (2009)
  [arXiv:0902.4050 [hep-ph]].


\bibitem{Iso:2012jn} 
  S.~Iso and Y.~Orikasa,
  TeV Scale B-L model with a flat Higgs potential at the Planck scale - in view of the hierarchy problem -,
  PTEP {\bf 2013}, 023B08 (2013)
  [arXiv:1210.2848 [hep-ph]].


\bibitem{Englert:2013gz} 
  C.~Englert, J.~Jaeckel, V.~V.~Khoze and M.~Spannowsky,
  Emergence of the Electroweak Scale through the Higgs Portal,
  JHEP {\bf 1304}, 060 (2013)
  [arXiv:1301.4224 [hep-ph]].


\bibitem{Abel:2013mya} 
  S.~Abel and A.~Mariotti,
  Novel Higgs Potentials from Gauge Mediation of Exact Scale Breaking,
  Phys.\ Rev.\ D {\bf 89}, no. 12, 125018 (2014)
  [arXiv:1312.5335 [hep-ph]].


\bibitem{Das:2015nwk} 
  A.~Das, N.~Okada and N.~Papapietro,
  Electroweak vacuum stability in classically conformal B-L extension of the Standard Model,
  Eur.\ Phys.\ J.\ C {\bf 77}, no. 2, 122 (2017)
  [arXiv:1509.01466 [hep-ph]].


\bibitem{Hashino:2015nxa} 
  K.~Hashino, S.~Kanemura and Y.~Orikasa,
  Discriminative phenomenological features of scale invariant models for electroweak symmetry breaking,
  Phys.\ Lett.\ B {\bf 752}, 217 (2016)
  [arXiv:1508.03245 [hep-ph]].


\bibitem{Kubo:2016kpb} 
  J.~Kubo and M.~Yamada,
  Scale genesis and gravitational wave in a classically scale invariant extension of the standard model,
  JCAP {\bf 1612}, no. 12, 001 (2016)
  [arXiv:1610.02241 [hep-ph]].


\bibitem{Kannike:2016wuy} 
  K.~Kannike, M.~Raidal, C.~Spethmann and H.~Veermäe,
  The evolving Planck mass in classically scale-invariant theories,
  JHEP {\bf 1704}, 026 (2017)
  [arXiv:1610.06571 [hep-ph]].


\bibitem{Ghilencea:2016dsl} 
  D.~M.~Ghilencea, Z.~Lalak and P.~Olszewski,
  Standard Model with spontaneously broken quantum scale invariance,
  Phys.\ Rev.\ D {\bf 96}, no. 5, 055034 (2017)
  [arXiv:1612.09120 [hep-ph]].


\bibitem{Das:2016zue} 
  A.~Das, S.~Oda, N.~Okada and D.~s.~Takahashi,
  Classically conformal U(1)′ extended standard model, electroweak vacuum stability, and LHC Run-2 bounds,
  Phys.\ Rev.\ D {\bf 93}, no. 11, 115038 (2016)
  [arXiv:1605.01157 [hep-ph]].


\bibitem{Arunasalam:2017ajm} 
  S.~Arunasalam, A.~Kobakhidze, C.~Lagger, S.~Liang and A.~Zhou,
  Low temperature electroweak phase transition in the Standard Model with hidden scale invariance,
  Phys.\ Lett.\ B {\bf 776}, 48 (2018)
  [arXiv:1709.10322 [hep-ph]].


\bibitem{Marzola:2017jzl} 
  L.~Marzola, A.~Racioppi and V.~Vaskonen,
  Phase transition and gravitational wave phenomenology of scalar conformal extensions of the Standard Model,
  Eur.\ Phys.\ J.\ C {\bf 77}, no. 7, 484 (2017)
  [arXiv:1704.01034 [hep-ph]].


\bibitem{Chataignier:2018kay} 
  L.~Chataignier, T.~Prokopec, M.~G.~Schmidt and B.~Świeżewska,
  Systematic analysis of radiative symmetry breaking in models with extended scalar sector,
  JHEP {\bf 1808}, 083 (2018)
  [arXiv:1805.09292 [hep-ph]].


\bibitem{Loebbert:2018xsd} 
  F.~Loebbert, J.~Miczajka and J.~Plefka,
  Consistent Conformal Extensions of the Standard Model,
  Phys.\ Rev.\ D {\bf 99}, no. 1, 015026 (2019)
  [arXiv:1805.09727 [hep-ph]].


\bibitem{AlexanderNunneley:2010nw} 
  L.~Alexander-Nunneley and A.~Pilaftsis,
  The Minimal Scale Invariant Extension of the Standard Model,
  JHEP {\bf 1009}, 021 (2010)
  [arXiv:1006.5916 [hep-ph]].


\bibitem{Masina:2013wja} 
  I.~Masina and M.~Quiros,
  On the Veltman Condition, the Hierarchy Problem and High-Scale Supersymmetry,
  Phys.\ Rev.\ D {\bf 88}, 093003 (2013)
  [arXiv:1308.1242 [hep-ph]].


\bibitem{Guo:2014bha} 
  J.~Guo and Z.~Kang,
  Higgs Naturalness and Dark Matter Stability by Scale Invariance,
  Nucl.\ Phys.\ B {\bf 898}, 415 (2015)
  [arXiv:1401.5609 [hep-ph]].


\bibitem{Chang:2007ki} 
  W.~F.~Chang, J.~N.~Ng and J.~M.~S.~Wu,
  Shadow Higgs from a scale-invariant hidden U(1)(s) model,
  Phys.\ Rev.\ D {\bf 75}, 115016 (2007)
  [hep-ph/0701254 [HEP-PH]].


\bibitem{Foot:2010av} 
  R.~Foot, A.~Kobakhidze and R.~R.~Volkas,
  Stable mass hierarchies and dark matter from hidden sectors in the scale-invariant standard model,
  Phys.\ Rev.\ D {\bf 82}, 035005 (2010)
  [arXiv:1006.0131 [hep-ph]].


\bibitem{Ishiwata:2011aa} 
  K.~Ishiwata,
  Dark Matter in Classically Scale-Invariant Two Singlets Standard Model,
  Phys.\ Lett.\ B {\bf 710}, 134 (2012)
  [arXiv:1112.2696 [hep-ph]].


\bibitem{Gabrielli:2013hma} 
  E.~Gabrielli, M.~Heikinheimo, K.~Kannike, A.~Racioppi, M.~Raidal and C.~Spethmann,
  Towards Completing the Standard Model: Vacuum Stability, EWSB and Dark Matter,
  Phys.\ Rev.\ D {\bf 89}, no. 1, 015017 (2014)
  [arXiv:1309.6632 [hep-ph]].


\bibitem{Endo:2015nba} 
  K.~Endo and K.~Ishiwata,
  Direct detection of singlet dark matter in classically scale-invariant standard model,
  Phys.\ Lett.\ B {\bf 749}, 583 (2015)
  [arXiv:1507.01739 [hep-ph]].


\bibitem{Wang:2015cda} 
  Z.~W.~Wang, T.~G.~Steele, T.~Hanif and R.~B.~Mann,
  Conformal Complex Singlet Extension of the Standard Model: Scenario for Dark Matter and a Second Higgs Boson,
  JHEP {\bf 1608}, 065 (2016)
  [arXiv:1510.04321 [hep-ph]].


\bibitem{Ghorbani:2015xvz} 
  K.~Ghorbani and H.~Ghorbani,
  Scalar Dark Matter in Scale Invariant Standard Model,
  JHEP {\bf 1604}, 024 (2016)
  [arXiv:1511.08432 [hep-ph]].


\bibitem{Plascencia:2015xwa} 
  A.~D.~Plascencia,
  Classical scale invariance in the inert doublet model,
  JHEP {\bf 1509}, 026 (2015)
  [arXiv:1507.04996 [hep-ph]].


\bibitem{Helmboldt:2016mpi} 
  A.~J.~Helmboldt, P.~Humbert, M.~Lindner and J.~Smirnov,
  Minimal conformal extensions of the Higgs sector,
  JHEP {\bf 1707}, 113 (2017)
  [arXiv:1603.03603 [hep-ph]].


\bibitem{Radovcic:2014rea} 
  S.~Benic and B.~Radovcic,
  Electroweak breaking and Dark Matter from the common scale,
  Phys.\ Lett.\ B {\bf 732}, 91 (2014)
  [arXiv:1401.8183 [hep-ph]].


\bibitem{Altmannshofer:2014vra} 
  W.~Altmannshofer, W.~A.~Bardeen, M.~Bauer, M.~Carena and J.~D.~Lykken,
  Light Dark Matter, Naturalness, and the Radiative Origin of the Electroweak Scale,
  JHEP {\bf 1501}, 032 (2015)
  [arXiv:1408.3429 [hep-ph]].


\bibitem{Benic:2014aga} 
  S.~Benic and B.~Radovcic,
  Majorana dark matter in a classically scale invariant model,
  JHEP {\bf 1501}, 143 (2015)
  [arXiv:1409.5776 [hep-ph]].


\bibitem{Ahriche:2015loa} 
  A.~Ahriche, K.~L.~McDonald and S.~Nasri,
  A Radiative Model for the Weak Scale and Neutrino Mass via Dark Matter,
  JHEP {\bf 1602}, 038 (2016)
  [arXiv:1508.02607 [hep-ph]].


\bibitem{Ahriche:2016ixu} 
  A.~Ahriche, A.~Manning, K.~L.~McDonald and S.~Nasri,
  Scale-Invariant Models with One-Loop Neutrino Mass and Dark Matter Candidates,
  Phys.\ Rev.\ D {\bf 94}, no. 5, 053005 (2016)
  [arXiv:1604.05995 [hep-ph]].


\bibitem{Oda:2017kwl} 
  S.~Oda, N.~Okada and D.~s.~Takahashi,
  Right-handed neutrino dark matter in the classically conformal U(1)′ extended standard model,
  Phys.\ Rev.\ D {\bf 96}, no. 9, 095032 (2017)
  [arXiv:1704.05023 [hep-ph]].


\bibitem{YaserAyazi:2018lrv} 
  S.~Yaser Ayazi and A.~Mohamadnejad,
  Scale-Invariant Two Component Dark Matter,
  Eur.\ Phys.\ J.\ C {\bf 79}, no. 2, 140 (2019)
  [arXiv:1808.08706 [hep-ph]].


\bibitem{Hambye:2013dgv} 
  T.~Hambye and A.~Strumia,
  Dynamical generation of the weak and Dark Matter scale,
  Phys.\ Rev.\ D {\bf 88}, 055022 (2013)
  [arXiv:1306.2329 [hep-ph]].


\bibitem{Carone:2013wla} 
  C.~D.~Carone and R.~Ramos,
  Classical scale-invariance, the electroweak scale and vector dark matter,
  Phys.\ Rev.\ D {\bf 88}, 055020 (2013)
  [arXiv:1307.8428 [hep-ph]].


\bibitem{Khoze:2014xha} 
  V.~V.~Khoze, C.~McCabe and G.~Ro,
  Higgs vacuum stability from the dark matter portal,
  JHEP {\bf 1408}, 026 (2014)
  [arXiv:1403.4953 [hep-ph]].


\bibitem{Karam:2015jta} 
  A.~Karam and K.~Tamvakis,
  Dark matter and neutrino masses from a scale-invariant multi-Higgs portal,
  Phys.\ Rev.\ D {\bf 92}, no. 7, 075010 (2015)
  [arXiv:1508.03031 [hep-ph]].


\bibitem{Karam:2016rsz} 
  A.~Karam and K.~Tamvakis,
  Dark Matter from a Classically Scale-Invariant $SU(3)_X$,
  Phys.\ Rev.\ D {\bf 94}, no. 5, 055004 (2016)
  [arXiv:1607.01001 [hep-ph]].


\bibitem{Khoze:2016zfi} 
  V.~V.~Khoze and A.~D.~Plascencia,
  Dark Matter and Leptogenesis Linked by Classical Scale Invariance,
  JHEP {\bf 1611}, 025 (2016)
  [arXiv:1605.06834 [hep-ph]].


\bibitem{Baldes:2018emh} 
  I.~Baldes and C.~Garcia-Cely,
  Strong gravitational radiation from a simple dark matter model,
  arXiv:1809.01198 [hep-ph].


\bibitem{Hambye:2008bq} 
  T.~Hambye,
  Hidden vector dark matter,
  JHEP {\bf 0901}, 028 (2009)
  [arXiv:0811.0172 [hep-ph]].


\bibitem{Hambye:2009fg} 
  T.~Hambye and M.~H.~G.~Tytgat,
  Confined hidden vector dark matter,
  Phys.\ Lett.\ B {\bf 683}, 39 (2010)
  [arXiv:0907.1007 [hep-ph]].


\bibitem{DiazCruz:2010dc} 
  J.~L.~Diaz-Cruz and E.~Ma,
  Neutral SU(2) Gauge Extension of the Standard Model and a Vector-Boson Dark-Matter Candidate,
  Phys.\ Lett.\ B {\bf 695}, 264 (2011)
  [arXiv:1007.2631 [hep-ph]].


\bibitem{Yu:2011by} 
  Z.~H.~Yu, J.~M.~Zheng, X.~J.~Bi, Z.~Li, D.~X.~Yao and H.~H.~Zhang,
  Constraining the interaction strength between dark matter and visible matter: II. scalar, vector and spin-3/2 dark matter,
  Nucl.\ Phys.\ B {\bf 860}, 115 (2012)
  [arXiv:1112.6052 [hep-ph]].


\bibitem{Farzan:2012hh} 
  Y.~Farzan and A.~R.~Akbarieh,
  VDM: A model for Vector Dark Matter,
  JCAP {\bf 1210}, 026 (2012)
  [arXiv:1207.4272 [hep-ph]].


\bibitem{Baek:2012se} 
  S.~Baek, P.~Ko, W.~I.~Park and E.~Senaha,
  Higgs Portal Vector Dark Matter : Revisited,
  JHEP {\bf 1305}, 036 (2013)
  [arXiv:1212.2131 [hep-ph]].


\bibitem{Baek:2013dwa} 
  S.~Baek, P.~Ko and W.~I.~Park,
  Hidden sector monopole, vector dark matter and dark radiation with Higgs portal,
  JCAP {\bf 1410}, no. 10, 067 (2014)
  [arXiv:1311.1035 [hep-ph]].


\bibitem{Davoudiasl:2013jma} 
  H.~Davoudiasl and I.~M.~Lewis,
  Dark Matter from Hidden Forces,
  Phys.\ Rev.\ D {\bf 89}, no. 5, 055026 (2014)
  [arXiv:1309.6640 [hep-ph]].


\bibitem{Fraser:2014yga} 
  S.~Fraser, E.~Ma and M.~Zakeri,
  $SU(2)_N$ model of vector dark matter with a leptonic connection,
  Int.\ J.\ Mod.\ Phys.\ A {\bf 30}, no. 03, 1550018 (2015)
  [arXiv:1409.1162 [hep-ph]].


\bibitem{Graham:2015rva} 
  P.~W.~Graham, J.~Mardon and S.~Rajendran,
  Vector Dark Matter from Inflationary Fluctuations,
  Phys.\ Rev.\ D {\bf 93}, no. 10, 103520 (2016)
  [arXiv:1504.02102 [hep-ph]].


\bibitem{DiChiara:2015bua} 
  S.~Di Chiara and K.~Tuominen,
  A minimal model for SU(N ) vector dark matter,
  JHEP {\bf 1511}, 188 (2015)
  [arXiv:1506.03285 [hep-ph]].


\bibitem{DiFranzo:2015nli} 
  A.~DiFranzo, P.~J.~Fox and T.~M.~P.~Tait,
  Vector Dark Matter through a Radiative Higgs Portal,
  JHEP {\bf 1604}, 135 (2016)
  [arXiv:1512.06853 [hep-ph]].


\bibitem{Cembranos:2016ugq} 
  J.~A.~R.~Cembranos, A.~L.~Maroto and S.~J.~Núñez Jareño,
  Perturbations of ultralight vector field dark matter,
  JHEP {\bf 1702}, 064 (2017)
  [arXiv:1611.03793 [astro-ph.CO]].


\bibitem{Choi:2017zww} 
  S.~M.~Choi, Y.~Hochberg, E.~Kuflik, H.~M.~Lee, Y.~Mambrini, H.~Murayama and M.~Pierre,
  Vector SIMP dark matter,
  JHEP {\bf 1710}, 162 (2017)
  [arXiv:1707.01434 [hep-ph]].


\bibitem{Duch:2017khv} 
  M.~Duch, B.~Grzadkowski and D.~Huang,
  Strongly self-interacting vector dark matter via freeze-in,
  JHEP {\bf 1801}, 020 (2018)
  [arXiv:1710.00320 [hep-ph]].


\bibitem{Ahmed:2017dbb} 
  A.~Ahmed, M.~Duch, B.~Grzadkowski and M.~Iglicki,
  Multi-Component Dark Matter: the vector and fermion case,
  Eur.\ Phys.\ J.\ C {\bf 78}, no. 11, 905 (2018)
  [arXiv:1710.01853 [hep-ph]].


\bibitem{Maru:2018ocf} 
  N.~Maru, N.~Okada and S.~Okada,
  $SU(2)_L$ doublet vector dark matter from gauge-Higgs unification,
  Phys.\ Rev.\ D {\bf 98}, no. 7, 075021 (2018)
  [arXiv:1803.01274 [hep-ph]].


\bibitem{Chakraborti:2018aae} 
  S.~Chakraborti, A.~Dutta Banik and R.~Islam,
  Probing Multicomponent Extension of Inert Doublet Model with a Vector Dark Matter,
  arXiv:1810.05595 [hep-ph].


\bibitem{Belyaev:2018xpf} 
  A.~Belyaev, G.~Cacciapaglia, J.~Mckay, D.~Marin and A.~R.~Zerwekh,
  Minimal Spin-one Isotriplet Dark Matter,
  arXiv:1808.10464 [hep-ph].


\bibitem{Saez:2018off} 
  B.~Díaz Sáez, F.~Rojas-Abatte and A.~R.~Zerwekh,
  Dark Matter from a Vector Field in the Fundamental Representation of $SU(2)_L$,
  arXiv:1810.06375 [hep-ph].


\bibitem{Hisano:2010yh} 
  J.~Hisano, K.~Ishiwata, N.~Nagata and M.~Yamanaka,
  Direct Detection of Vector Dark Matter,
  Prog.\ Theor.\ Phys.\  {\bf 126}, 435 (2011)
  [arXiv:1012.5455 [hep-ph]].


\bibitem{Baek:2014goa} 
  S.~Baek, P.~Ko, W.~I.~Park and Y.~Tang,
  Indirect and direct signatures of Higgs portal decaying vector dark matter for positron excess in cosmic rays,
  JCAP {\bf 1406}, 046 (2014)
  [arXiv:1402.2115 [hep-ph]].


\bibitem{Yu:2014pra} 
  J.~H.~Yu,
  Vector Fermion-Portal Dark Matter: Direct Detection and Galactic Center Gamma-Ray Excess,
  Phys.\ Rev.\ D {\bf 90}, no. 9, 095010 (2014)
  [arXiv:1409.3227 [hep-ph]].


\bibitem{Chen:2014cbt} 
  C.~R.~Chen, Y.~K.~Chu and H.~C.~Tsai,
  An Elusive Vector Dark Matter,
  Phys.\ Lett.\ B {\bf 741}, 205 (2015)
  [arXiv:1410.0918 [hep-ph]].


\bibitem{Yang:2016zaz} 
  Q.~Yang and H.~Di,
  Vector Dark Matter Detection using the Quantum Jump of Atoms,
  Phys.\ Lett.\ B {\bf 780}, 622 (2018)
  [arXiv:1606.01492 [hep-ph]].


\bibitem{Chen:2016rae} 
  C.~R.~Chen and M.~J.~Li,
  New LUX result constrains exotic quark mediators with the vector dark matter,
  Int.\ J.\ Mod.\ Phys.\ A {\bf 31}, no. 36, 1650200 (2016)
  [arXiv:1609.07583 [hep-ph]].


\bibitem{Catena:2018uae} 
  R.~Catena, K.~Fridell and V.~Zema,
  Direct detection of fermionic and vector dark matter with polarised targets,
  JCAP {\bf 1811}, no. 11, 018 (2018)
  [arXiv:1810.01515 [hep-ph]].


\bibitem{Yepes:2018zkk} 
  J.~Yepes,
  Top partners tackling vector dark matter,
  arXiv:1811.06059 [hep-ph].


\bibitem{Arina:2009uq} 
  C.~Arina, T.~Hambye, A.~Ibarra and C.~Weniger,
  Intense Gamma-Ray Lines from Hidden Vector Dark Matter Decay,
  JCAP {\bf 1003}, 024 (2010)
  [arXiv:0912.4496 [hep-ph]].


\bibitem{Farzan:2012kk} 
  Y.~Farzan and A.~R.~Akbarieh,
  Natural explanation for 130 GeV photon line within vector boson dark matter model,
  Phys.\ Lett.\ B {\bf 724}, 84 (2013)
  [arXiv:1211.4685 [hep-ph]].


\bibitem{Choi:2013eua} 
  K.~Y.~Choi, H.~M.~Lee and O.~Seto,
  Vector Higgs-portal dark matter and Fermi-LAT gamma ray line,
  Phys.\ Rev.\ D {\bf 87}, no. 12, 123541 (2013)
  [arXiv:1304.0966 [hep-ph]].


\bibitem{Ko:2014gha} 
  P.~Ko, W.~I.~Park and Y.~Tang,
  Higgs portal vector dark matter for $\mathinner{\mathrm{GeV}}$ scale $\gamma$-ray excess from galactic center,
  JCAP {\bf 1409}, 013 (2014)
  [arXiv:1404.5257 [hep-ph]].


\bibitem{Farzan:2014foo} 
  Y.~Farzan and A.~R.~Akbarieh,
  Decaying Vector Dark Matter as an Explanation for the 3.5 keV Line from Galaxy Clusters,
  JCAP {\bf 1411}, no. 11, 015 (2014)
  [arXiv:1408.2950 [hep-ph]].


\bibitem{Bambhaniya:2016cpr} 
  G.~Bambhaniya, J.~Kumar, D.~Marfatia, A.~C.~Nayak and G.~Tomar,
  Vector dark matter annihilation with internal bremsstrahlung,
  Phys.\ Lett.\ B {\bf 766}, 177 (2017)
  [arXiv:1609.05369 [hep-ph]].


\bibitem{Chen:2017tva} 
  C.~H.~Chen, C.~W.~Chiang and T.~Nomura,
  Explaining the DAMPE $e^+ e^-$ excess using the Higgs triplet model with a vector dark matter,
  Phys.\ Rev.\ D {\bf 97}, no. 6, 061302 (2018)
  [arXiv:1712.00793 [hep-ph]].


\bibitem{Yang:2018fje} 
  K.~C.~Yang,
  Hidden Higgs portal vector dark matter for the Galactic center gamma-ray excess from the two-step cascade annihilation, and muon g − 2,
  JHEP {\bf 1808}, 099 (2018)
  [arXiv:1806.05663 [hep-ph]].


\bibitem{Lebedev:2011iq} 
  O.~Lebedev, H.~M.~Lee and Y.~Mambrini,
  Vector Higgs-portal dark matter and the invisible Higgs,
  Phys.\ Lett.\ B {\bf 707}, 570 (2012)
  [arXiv:1111.4482 [hep-ph]].


\bibitem{Duch:2015jta} 
  M.~Duch, B.~Grzadkowski and M.~McGarrie,
  A stable Higgs portal with vector dark matter,
  JHEP {\bf 1509}, 162 (2015)
  [arXiv:1506.08805 [hep-ph]].


\bibitem{Chen:2015dea} 
  C.~H.~Chen and T.~Nomura,
  Searching for vector dark matter via Higgs portal at the LHC,
  Phys.\ Rev.\ D {\bf 93}, no. 7, 074019 (2016)
  [arXiv:1507.00886 [hep-ph]].


\bibitem{Kumar:2015wya} 
  J.~Kumar, D.~Marfatia and D.~Yaylali,
  Vector dark matter at the LHC,
  Phys.\ Rev.\ D {\bf 92}, no. 9, 095027 (2015)
  [arXiv:1508.04466 [hep-ph]].


\bibitem{Barman:2017yzr} 
  B.~Barman, S.~Bhattacharya, S.~K.~Patra and J.~Chakrabortty,
  Non-Abelian Vector Boson Dark Matter, its Unified Route and signatures at the LHC,
  JCAP {\bf 1712}, no. 12, 021 (2017)
  [arXiv:1704.04945 [hep-ph]].


\bibitem{Farzinnia:2013pga} 
  A.~Farzinnia, H.~J.~He and J.~Ren,
  Natural Electroweak Symmetry Breaking from Scale Invariant Higgs Mechanism,
  Phys.\ Lett.\ B {\bf 727}, 141 (2013)
  [arXiv:1308.0295 [hep-ph]].


\bibitem{Farzinnia:2014xia} 
  A.~Farzinnia and J.~Ren,
  Higgs Partner Searches and Dark Matter Phenomenology in a Classically Scale Invariant Higgs Boson Sector,
  Phys.\ Rev.\ D {\bf 90}, no. 1, 015019 (2014)
  [arXiv:1405.0498 [hep-ph]].


\bibitem{Sakharov:1967dj} 
  A.~D.~Sakharov,
  Violation of CP Invariance, C asymmetry, and baryon asymmetry of the universe,
  Pisma Zh.\ Eksp.\ Teor.\ Fiz.\  {\bf 5}, 32 (1967)
  [JETP Lett.\  {\bf 5}, 24 (1967)]
  [Sov.\ Phys.\ Usp.\  {\bf 34}, no. 5, 392 (1991)]
  [Usp.\ Fiz.\ Nauk {\bf 161}, no. 5, 61 (1991)].


\bibitem{Carrington:1991hz} 
  M.~E.~Carrington,
  The Effective potential at finite temperature in the Standard Model,
  Phys.\ Rev.\ D {\bf 45}, 2933 (1992).


\bibitem{Anderson:1991zb} 
  G.~W.~Anderson and L.~J.~Hall,
  The Electroweak phase transition and baryogenesis,
  Phys.\ Rev.\ D {\bf 45}, 2685 (1992).


\bibitem{Arnold:1992fb} 
  P.~B.~Arnold,
  Phase transition temperatures at next-to-leading order,
  Phys.\ Rev.\ D {\bf 46}, 2628 (1992)
  [hep-ph/9204228].


\bibitem{Arnold:1992rz} 
  P.~B.~Arnold and O.~Espinosa,
  The Effective potential and first order phase transitions: Beyond leading-order,
  Phys.\ Rev.\ D {\bf 47}, 3546 (1993)
  Erratum: [Phys.\ Rev.\ D {\bf 50}, 6662 (1994)]
  [hep-ph/9212235].


\bibitem{Dine:1992wr} 
  M.~Dine, R.~G.~Leigh, P.~Y.~Huet, A.~D.~Linde and D.~A.~Linde,
  Towards the theory of the electroweak phase transition,
  Phys.\ Rev.\ D {\bf 46}, 550 (1992)
  [hep-ph/9203203].


\bibitem{Chatrchyan:2012xdj} 
  S.~Chatrchyan {\it et al.} [CMS Collaboration],
  Observation of a new boson at a mass of 125 GeV with the CMS experiment at the LHC,
  Phys.\ Lett.\ B {\bf 716}, 30 (2012)
  [arXiv:1207.7235 [hep-ex]].


\bibitem{Aad:2012tfa} 
  G.~Aad {\it et al.} [ATLAS Collaboration],
  Observation of a new particle in the search for the Standard Model Higgs boson with the ATLAS detector at the LHC,
  Phys.\ Lett.\ B {\bf 716}, 1 (2012)
  [arXiv:1207.7214 [hep-ex]].


\bibitem{Dimopoulos:1990ai} 
  S.~Dimopoulos, R.~Esmailzadeh, L.~J.~Hall and N.~Tetradis,
  Electroweak phase transition and dark matter abundance,
  Phys.\ Lett.\ B {\bf 247}, 601 (1990).


\bibitem{Chung:2011it} 
  D.~J.~H.~Chung and A.~J.~Long,
  Cosmological Constant, Dark Matter, and Electroweak Phase Transition,
  Phys.\ Rev.\ D {\bf 84}, 103513 (2011)
  [arXiv:1108.5193 [astro-ph.CO]].


\bibitem{Carena:2011jy} 
  M.~Carena, N.~R.~Shah and C.~E.~M.~Wagner,
  Light Dark Matter and the Electroweak Phase Transition in the NMSSM,
  Phys.\ Rev.\ D {\bf 85}, 036003 (2012)
  [arXiv:1110.4378 [hep-ph]].


\bibitem{Chowdhury:2011ga} 
  T.~A.~Chowdhury, M.~Nemevsek, G.~Senjanovic and Y.~Zhang,
  Dark Matter as the Trigger of Strong Electroweak Phase Transition,
  JCAP {\bf 1202}, 029 (2012)
  [arXiv:1110.5334 [hep-ph]].


\bibitem{Ahriche:2012ei} 
  A.~Ahriche and S.~Nasri,
  Light Dark Matter, Light Higgs and the Electroweak Phase Transition,
  Phys.\ Rev.\ D {\bf 85}, 093007 (2012)
  [arXiv:1201.4614 [hep-ph]].


\bibitem{Borah:2012pu} 
  D.~Borah and J.~M.~Cline,
  Inert Doublet Dark Matter with Strong Electroweak Phase Transition,
  Phys.\ Rev.\ D {\bf 86}, 055001 (2012)
  [arXiv:1204.4722 [hep-ph]].


\bibitem{Gil:2012ya} 
  G.~Gil, P.~Chankowski and M.~Krawczyk,
  Inert Dark Matter and Strong Electroweak Phase Transition,
  Phys.\ Lett.\ B {\bf 717}, 396 (2012)
  [arXiv:1207.0084 [hep-ph]].


\bibitem{Falkowski:2012fb} 
  A.~Falkowski and J.~M.~No,
  Non-thermal Dark Matter Production from the Electroweak Phase Transition: Multi-TeV WIMPs and 'Baby-Zillas',
  JHEP {\bf 1302}, 034 (2013)
  [arXiv:1211.5615 [hep-ph]].


\bibitem{Cline:2013bln} 
  J.~M.~Cline and K.~Kainulainen,
  Improved Electroweak Phase Transition with Subdominant Inert Doublet Dark Matter,
  Phys.\ Rev.\ D {\bf 87}, no. 7, 071701 (2013)
  [arXiv:1302.2614 [hep-ph]].


\bibitem{Ahriche:2013zwa} 
  A.~Ahriche and S.~Nasri,
  Dark matter and strong electroweak phase transition in a radiative neutrino mass model,
  JCAP {\bf 1307}, 035 (2013)
  [arXiv:1304.2055 [hep-ph]].


\bibitem{Fairbairn:2013uta} 
  M.~Fairbairn and R.~Hogan,
  Singlet Fermionic Dark Matter and the Electroweak Phase Transition,
  JHEP {\bf 1309}, 022 (2013)
  [arXiv:1305.3452 [hep-ph]].


\bibitem{AbdusSalam:2013eya} 
  S.~S.~AbdusSalam and T.~A.~Chowdhury,
  Scalar Representations in the Light of Electroweak Phase Transition and Cold Dark Matter Phenomenology,
  JCAP {\bf 1405}, 026 (2014)
  [arXiv:1310.8152 [hep-ph]].


\bibitem{Chowdhury:2014tpa} 
  T.~A.~Chowdhury,
  A Possible Link between the Electroweak Phase Transition and the Dark Matter of the Universe, Ph.D. Thesis, SISSA, Trieste, Italy (2014).


\bibitem{Chao:2014ina} 
  W.~Chao,
  First order electroweak phase transition triggered by the Higgs portal vector dark matter,
  Phys.\ Rev.\ D {\bf 92}, no. 1, 015025 (2015)
  [arXiv:1412.3823 [hep-ph]].


\bibitem{Chao:2017vrq} 
  W.~Chao, H.~K.~Guo and J.~Shu,
  Gravitational Wave Signals of Electroweak Phase Transition Triggered by Dark Matter,
  JCAP {\bf 1709}, no. 09, 009 (2017)
  [arXiv:1702.02698 [hep-ph]].


\bibitem{Gu:2017rzz} 
  P.~H.~Gu,
  Cosmic matter from dark electroweak phase transition with neutrino mass generation,
  Phys.\ Rev.\ D {\bf 96}, no. 5, 055038 (2017)
  [arXiv:1705.05189 [hep-ph]].


\bibitem{Liu:2017gfg} 
  X.~Liu and L.~Bian,
  Dark matter and electroweak phase transition in the mixed scalar dark matter model,
  Phys.\ Rev.\ D {\bf 97}, no. 5, 055028 (2018)
  [arXiv:1706.06042 [hep-ph]].


\bibitem{Ghorbani:2017lyk} 
  P.~H.~Ghorbani,
  Electroweak phase transition in the scale invariant standard model,
  Phys.\ Rev.\ D {\bf 98}, no. 11, 115016 (2018)
  [arXiv:1711.11541 [hep-ph]].


\bibitem{Shajiee:2018jdq} 
  V.~R.~Shajiee and A.~Tofighi,
  Electroweak Phase Transition, Gravitational Waves and Dark Matter in Two Scalar Singlet Extension of The Standard Model,
  arXiv:1811.09807 [hep-ph].


\bibitem{Aghanim:2018eyx} 
  N.~Aghanim {\it et al.} [Planck Collaboration],
  Planck 2018 results. VI. Cosmological parameters,
  arXiv:1807.06209 [astro-ph.CO].


\bibitem{Hinshaw:2012aka} 
  G.~Hinshaw {\it et al.} [WMAP Collaboration],
  Nine-Year Wilkinson Microwave Anisotropy Probe (WMAP) Observations: Cosmological Parameter Results,
  Astrophys.\ J.\ Suppl.\  {\bf 208}, 19 (2013)
  [arXiv:1212.5226 [astro-ph.CO]].


\bibitem{Cui:2017nnn} 
  X.~Cui {\it et al.} [PandaX-II Collaboration],
  Dark Matter Results From 54-Ton-Day Exposure of PandaX-II Experiment,
  Phys.\ Rev.\ Lett.\  {\bf 119}, no. 18, 181302 (2017)
  [arXiv:1708.06917 [astro-ph.CO]].


\bibitem{Gildener:1976ih} 
  E.~Gildener and S.~Weinberg,
  Symmetry Breaking and Scalar Bosons,
  Phys.\ Rev.\ D {\bf 13}, 3333 (1976).


\bibitem{Barducci:2016pcb} 
  D.~Barducci, G.~Belanger, J.~Bernon, F.~Boudjema, J.~Da Silva, S.~Kraml, U.~Laa and A.~Pukhov,
  Collider limits on new physics within micrOMEGAs-4.3,
  Comput.\ Phys.\ Commun.\  {\bf 222}, 327 (2018)
  [arXiv:1606.03834 [hep-ph]].


\bibitem{Semenov:2014rea} 
  A.~Semenov,
  LanHEP — A package for automatic generation of Feynman rules from the Lagrangian. Version 3.2,
  Comput.\ Phys.\ Commun.\  {\bf 201}, 167 (2016)
  [arXiv:1412.5016 [physics.comp-ph]].


\bibitem{Akerib:2016vxi} 
  D.~S.~Akerib {\it et al.} [LUX Collaboration],
  Results from a search for dark matter in the complete LUX exposure,
  Phys.\ Rev.\ Lett.\  {\bf 118}, no. 2, 021303 (2017)
  [arXiv:1608.07648 [astro-ph.CO]].


\bibitem{Aprile:2017iyp} 
  E.~Aprile {\it et al.} [XENON Collaboration],
  First Dark Matter Search Results from the XENON1T Experiment,
  Phys.\ Rev.\ Lett.\  {\bf 119}, no. 18, 181301 (2017)
  [arXiv:1705.06655 [astro-ph.CO]].


\bibitem{Billard:2013qya} 
  J.~Billard, L.~Strigari and E.~Figueroa-Feliciano,
  Implication of neutrino backgrounds on the reach of next generation dark matter direct detection experiments,
  Phys.\ Rev.\ D {\bf 89}, no. 2, 023524 (2014)
  [arXiv:1307.5458 [hep-ph]].


\bibitem{Cabrera:1984rr} 
  B.~Cabrera, L.~M.~Krauss and F.~Wilczek,
  Bolometric Detection of Neutrinos,
  Phys.\ Rev.\ Lett.\  {\bf 55}, 25 (1985).


\bibitem{Monroe:2007xp} 
  J.~Monroe and P.~Fisher,
  Neutrino Backgrounds to Dark Matter Searches,
  Phys.\ Rev.\ D {\bf 76}, 033007 (2007)
  [arXiv:0706.3019 [astro-ph]].


\bibitem{Strigari:2009bq} 
  L.~E.~Strigari,
  Neutrino Coherent Scattering Rates at Direct Dark Matter Detectors,
  New J.\ Phys.\  {\bf 11}, 105011 (2009)
  [arXiv:0903.3630 [astro-ph.CO]].


\bibitem{Gutlein:2010tq} 
  A.~Gutlein {\it et al.},
  Solar and atmospheric neutrinos: Background sources for the direct dark matter search,
  Astropart.\ Phys.\  {\bf 34}, 90 (2010)
  [arXiv:1003.5530 [hep-ph]].


\bibitem{Harnik:2012ni} 
  R.~Harnik, J.~Kopp and P.~A.~N.~Machado,
  Exploring nu Signals in Dark Matter Detectors,
  JCAP {\bf 1207}, 026 (2012)
  [arXiv:1202.6073 [hep-ph]].


\bibitem{Ruppin:2014bra} 
  F.~Ruppin, J.~Billard, E.~Figueroa-Feliciano and L.~Strigari,
  Complementarity of dark matter detectors in light of the neutrino background,
  Phys.\ Rev.\ D {\bf 90}, no. 8, 083510 (2014)
  [arXiv:1408.3581 [hep-ph]].


\bibitem{Davis:2014ama} 
  J.~H.~Davis,
  Dark Matter vs. Neutrinos: The effect of astrophysical uncertainties and timing information on the neutrino floor,
  JCAP {\bf 1503}, 012 (2015)
  [arXiv:1412.1475 [hep-ph]].


\bibitem{Dutta:2015vwa} 
  B.~Dutta, R.~Mahapatra, L.~E.~Strigari and J.~W.~Walker,
  Sensitivity to $Z$-prime and nonstandard neutrino interactions from ultralow threshold neutrino-nucleus coherent scattering,
  Phys.\ Rev.\ D {\bf 93}, no. 1, 013015 (2016)
  [arXiv:1508.07981 [hep-ph]].


\bibitem{Dent:2016iht} 
  J.~B.~Dent, B.~Dutta, J.~L.~Newstead and L.~E.~Strigari,
  Effective field theory treatment of the neutrino background in direct dark matter detection experiments,
  Phys.\ Rev.\ D {\bf 93}, no. 7, 075018 (2016)
  [arXiv:1602.05300 [hep-ph]].


\bibitem{Ng:2017aur} 
  K.~C.~Y.~Ng, J.~F.~Beacom, A.~H.~G.~Peter and C.~Rott,
  Solar Atmospheric Neutrinos: A New Neutrino Floor for Dark Matter Searches,
  Phys.\ Rev.\ D {\bf 96}, no. 10, 103006 (2017)
  [arXiv:1703.10280 [astro-ph.HE]].


\bibitem{Aprile:2015uzo} 
  E.~Aprile {\it et al.} [XENON Collaboration],
  Physics reach of the XENON1T dark matter experiment,
  JCAP {\bf 1604}, no. 04, 027 (2016)
  [arXiv:1512.07501 [physics.ins-det]].


\bibitem{Szydagis:2016few} 
  M.~Szydagis [LUX and LZ Collaborations],
  The Present and Future of Searching for Dark Matter with LUX and LZ,
  PoS ICHEP {\bf 2016}, 220 (2016)
  [arXiv:1611.05525 [astro-ph.CO]].


\bibitem{Aalbers:2016jon} 
  J.~Aalbers {\it et al.} [DARWIN Collaboration],
  DARWIN: towards the ultimate dark matter detector,
  JCAP {\bf 1611}, 017 (2016)
  [arXiv:1606.07001 [astro-ph.IM]].


\bibitem{DeSimone:2011ek} 
  A.~De Simone, G.~Nardini, M.~Quiros and A.~Riotto,
  Magnetic Fields at First Order Phase Transition: A Threat to Electroweak Baryogenesis,
  JCAP {\bf 1110}, 030 (2011)
  [arXiv:1107.4317 [hep-ph]].


\bibitem{Shaposhnikov:1987tw} 
  M.~E.~Shaposhnikov,
  Baryon Asymmetry of the Universe in Standard Electroweak Theory,
  Nucl.\ Phys.\ B {\bf 287}, 757 (1987).


\bibitem{Shaposhnikov:1986jp} 
  M.~E.~Shaposhnikov,
  Possible Appearance of the Baryon Asymmetry of the Universe in an Electroweak Theory,
  JETP Lett.\  {\bf 44}, 465 (1986)
  [Pisma Zh.\ Eksp.\ Teor.\ Fiz.\  {\bf 44}, 364 (1986)].


\bibitem{Dolan:1973qd} 
  L.~Dolan and R.~Jackiw,
  Symmetry Behavior at Finite Temperature,
  Phys.\ Rev.\ D {\bf 9}, 3320 (1974).


\bibitem{Conrad:2014tla} 
  J.~Conrad,
  Indirect Detection of WIMP Dark Matter: a compact review,
  arXiv:1411.1925 [hep-ph].


\bibitem{Ambrosi:2017wek} 
  G.~Ambrosi {\it et al.} [DAMPE Collaboration],
  Direct detection of a break in the teraelectronvolt cosmic-ray spectrum of electrons and positrons,
  Nature {\bf 552}, 63 (2017)
  [arXiv:1711.10981 [astro-ph.HE]].


\bibitem{Athron:2017drj} 
  P.~Athron, C.~Balazs, A.~Fowlie and Y.~Zhang,
  Model-independent analysis of the DAMPE excess,
  JHEP {\bf 1802}, 121 (2018)
  [arXiv:1711.11376 [hep-ph]].


\bibitem{Fowlie:2017fya} 
  A.~Fowlie,
  DAMPE squib? Significance of the 1.4 TeV DAMPE excess,
  Phys.\ Lett.\ B {\bf 780}, 181 (2018)
  [arXiv:1712.05089 [hep-ph]].


\bibitem{Yuan:2017ysv} 
  Q.~Yuan {\it et al.},
  Interpretations of the DAMPE electron data,
  arXiv:1711.10989 [astro-ph.HE].

\end{thebibliography}
\end{document}